\documentclass[aps,reprint,floatfix,showpacs,superscriptaddress]{revtex4-2}
\usepackage{natbib}
\usepackage{comment}
\usepackage{amsfonts,amsmath,amssymb}
\usepackage{bm}
\usepackage{pgf}
\usepackage{comment}
\newcommand{\antifrac}{\phi}

\begin{document}
\title{Cluster percolation in the three-dimensional $\pm J$ random-bond Ising model}
\author{L. M\"unster}
\affiliation{Institut f\"ur Physik, Technische Universit\"at Chemnitz, 09107 Chemnitz, Germany}
\author{M. Weigel}
\affiliation{Institut f\"ur Physik, Technische Universit\"at Chemnitz, 09107 Chemnitz, Germany}
\affiliation{Physics Department, Emory University, Atlanta, GA, U.S.A.}
\date{\today}

\begin{abstract}
  Based on extensive parallel-tempering Monte Carlo simulations, we investigate the
  relationship between cluster percolation and equilibrium ordering phenomena in the
  three-dimensional $\pm J$ random-bond Ising model as one varies the fraction of
  antiferromagnetic bonds. We consider a range of cluster definitions, most of which
  are constructed in the space of overlaps between two independent real replicas of
  the system. In the pure ferromagnet that is contained as a limiting case in the
  class of problems considered, the relevant percolation point coincides with the
  thermodynamic ordering transition. For the disordered ferromagnet encountered first
  on introducing antiferromagnetic bonds and the adjacent spin-glass phase of strong
  disorder this connection is altered, and one finds a percolation transition above
  the thermodynamic ordering point that is accompanied by the appearance of
  \emph{two} percolating clusters of equal density. Only at the lower (disordered)
  ferromagnetic or spin-glass transition points the densities of these two clusters
  start to diverge, thus providing a percolation signature of these thermodynamic
  transitions. We compare the scaling behavior at this secondary percolation
  transition with the thermodynamic behavior at the corresponding ferromagnetic and
  spin-glass phase transitions.
\end{abstract}

\maketitle
\section{Introduction}
\label{sec:introduction}

Cluster-based approaches provide elegant and versatile descriptions of the phenomena
at continuous thermodynamic phase
transitions~\cite{fisher:67,luijten:06,ConiglioFierro2021CorrelatedPercolation}. While
the list of applications of such approaches is long, and new systems with geometric
descriptions keep being added (see, e.g.,
Ref.~\cite{Almeida2023CriticalPercolationInTheOrderingKineticsOfTwistedNematicPhases}),
this is particularly true for unfrustrated lattice spin
models~\cite{ConiglioFierro2021CorrelatedPercolation,Grimmett2004TheRandomClusterModel}. A
prototypical example is the usage of Fortuin--Kasteleyn--Coniglio--Klein (FKCK)
clusters in the Ising
ferromagnet~\cite{FortuinKasteleyn1972OnTheRandomClusterModel1,ConiglioKlein1980ClustersAndCriticalDroplets}. The
construction of these clusters allows for a mapping between the
paramagnetic--ferromagnetic phase transition and the percolation transition of FKCK
clusters, where the density of the largest cluster is equal to the
magnetization. Furthermore, the corresponding Swendsen--Wang cluster Monte Carlo
dynamics~\cite{SwendsenWang1987NonuniversalCriticalDynamicsInMonteCarloSimulations,Wolff1988CollectiveMCUpdatingForSpinSystems},
which is based on FKCK clusters, drastically reduces the effect of critical slowing
down in comparison to local algorithms and thus provides a powerful tool for
simulations~\cite{Janke1998NonlocalMonteCarloAlgorithmsForStatisticalPhysicsApplications,Wolff1989ComparisonBetweenClusterMCAlgorithmsInTheIsingModel}.

For frustrated systems such as spin
glasses~\cite{BinderYoung1986SpinGlasses,NewmanStein2013SpinGlassesAndComplexity,Nishimori2001StatisticalPhysicsOfSpinGlassesAndInformationProcessing,CharbonneauEtAl2023SpinGlassesAndFarBeyond},
a similarly close connection between the percolation of spin clusters and the thermal
phase transition has not in general been established. In fact, it has been shown that
in such systems spin clusters and even FKCK clusters percolate at high temperatures
far above the ordering
transition~\cite{DeArcangelis1991PercolationTransitionInSpinGlasses,ConiglioEtAl1991ClusterApproachToSpinGlassesAndTheFrustratedPercolationProblem,DeSantisGandolfi1999BondPercolationInFrustratedSystems,FajenHartmannYoung2020percolationPercolationOfFortuinKasteleynClustersForTheRandomBondIsingModel,MuensterWeigel2023ClusterPercolationInTheTwoDimensionalIsingSpinGlass}. While
in the pure ferromagnet and even in unfrustrated disordered magnets the spin-spin
correlation function follows from the probability $\gamma_{\bm{x}\bm{y}}$ for spins at $\bm{x}$
and $\bm{y}$ to be in the same cluster,
\begin{equation}
  |\langle s_{\bm{x}} s_{\bm{y}}\rangle | = \langle \gamma_{\bm{x}\bm{y}} \rangle,
  \label{eq:Ising-FKCK-correspondence}
\end{equation}
in the presence of frustration this identity is weakened to an inequality, viz.
\begin{equation}
  |\langle s_{\bm{x}} s_{\bm{y}}\rangle | \le \langle \gamma_{\bm{x}\bm{y}} \rangle.
  \label{eq:broken-Ising-FKCK-correspondence}
\end{equation}
Consequently, the direct relation between the density of the largest cluster and the
order parameter in the ferromagnet is lost for the case of frustrated systems, where FKCK
cluster percolation no longer implies the presence of long-range
order~\cite{CataudellaEtAl1994CriticalClustersAndEfficientDynamicsForFrustratedSpinModels} (but see Ref.~\cite{miranda:25}
for a possible generalization). For
spin glasses, long-range ferromagnetic order is absent and, instead, there is
spin-glass ordering evidenced by the appearance of a non-zero overlap between two
independent replicas~\cite{BinderYoung1986SpinGlasses}. It is hence natural to
consider cluster definitions derived from multiple
replicas~\cite{SwendsenWang1986ReplicaMonteCarloSimulationOfSpinGlasses,AkritidisEtAl2023GeometricClustersInTheOverlapOfTheIsingModel,MuensterWeigel2023ClusterPercolationInTheTwoDimensionalIsingSpinGlass,MuensterWeigel2024SpinGlassesAndPercolation}
to describe such transitions. Chayes, Machta, and
Redner~\cite{CMR1998GraphicalRepresentationsForIsingSystemsInExternalFields} as well
as J\"org~\cite{Joerg2005ClusterMonteCarloAlgorithmsForDilutedSpinGlasses} (CMRJ)
have provided a cluster definition in this spirit.

CMRJ clusters have been studied in the Sherrington--Kirkpatrick (SK)
model~\cite{sherrington:75}, representing the mean-field limit of spin glasses, where
it was shown that the density \emph{difference} of the two largest clusters is equal
to the overlap, becoming non-zero at the spin-glass
transition~\cite{MachtaNewmanStein2008ThePercolationSignatureOfTheSpinGlassTransition,MachtaNewmanStein2009APercolationTheoreticApproachToSGPhaseTransitions}. In
two dimensions, on the other hand, CMRJ clusters asymptotically do not percolate at
non-zero temperature, in agreement with the zero-temperature spin-glass transition
there~\cite{MuensterWeigel2023ClusterPercolationInTheTwoDimensionalIsingSpinGlass}. These
CMRJ clusters as well as a number of related cluster definitions such as the ones due
to Houdayer~\cite{Houdayer2001ClusterMonteCarloAlgorithmFor2DimensionalSpinGlasses}
and the two-replica FKCK
construction~\cite{NewmanStein2007ShortRangeSpinGlassesResultsAndSpeculations} have
also found numerous numerical applications in Monte Carlo
simulations~\cite{SwendsenWang1986ReplicaMonteCarloSimulationOfSpinGlasses,
  WangSwendsen1988LowTemperaturePropertiesOfThePMJIsingSpinGlassInTwoDimensions,
  CMR1998GraphicalRepresentationsAndClusterAlgorithmsForCriticalPointsWithFields,
  MachtaNewmanChayes2000ReplicaExchangeAlgorithmAndResultsForTheThreeDimensionalRandomFieldIsingModel,
  Houdayer2001ClusterMonteCarloAlgorithmFor2DimensionalSpinGlasses,
  WangSwendsen2005ReplicaMonteCarloSimulationRevisited,
  Joerg2006CriticalBehaviorOfThe3DBondDilutedIsingSpinGlass,
  Hasenbuch2013ThermodynamicCasimirForcesBetweenASphereAndAPlateMonteCarloSimulationOfASpinModel,
  ZhuEtAl2015EfficientClusterAlgorithmForSpinGlassesInAnySpaceDimension,
  Hasenbusch2016VarianceReducedEstimatorOfTheConnectedTwoPointFunctionInThePresenceOfABrokenZ2Symmetry}.

In the present work we explore the behavior of different cluster types for the
crucial case of three dimensions that is physical and where a finite-temperature
spin-glass phase is present. To arrive at a fuller picture we consider the general
case of the $\pm J$ random-bond Ising model with different fractions of
antiferromagnetic bonds, $0 \leq \antifrac \leq 0.5$.  Specifically, the model is
examined in detail for $\antifrac=0$, representing a pure ferromagnet;
$\antifrac=0.125$, corresponding to a disordered and frustrated ferromagnet; and
$\antifrac=0.5$, which describes a spin glass. This approach provides insight into
the relationship between thermal phase transitions and cluster percolation at
different levels of frustration. Using extensive Markov chain Monte Carlo simulations
together with suitable finite-size scaling (FSS) analyses, we study percolation
properties of the considered cluster types. In particular, we are focused on the
question of when percolating clusters first appear and when density differences
between percolating clusters set in. We find that these events coincide with the
occurrence of the relevant ordering transitions and, where appropriate, we consider the relation of
the corresponding percolation critical exponents to the exponents of the
associated thermal transitions.

The rest of this paper is organized as follows. In
Sec.~\ref{sec:model_phase_diagram_and_cluster_definitions} we introduce the $\pm J$
random-bond Ising model and the associated phase diagram. We then move on to discuss
the relevant cluster definitions and some of their basic properties.
Section~\ref{sec:observables_and_finite_size_scaling} provides a summary of the
considered observables together with the expected FSS behavior. The following three
sections present our results for the cases of the pure ferromagnet
(Sec.~\ref{sec:clusters_in_the_pure_ferromagnet}), which we studied as a reference,
the intermediate case of the disordered and frustrated ferromagnet
(Sec.~\ref{sec:clusters_in_the_frustrated_disordered_ferromagnet}), and the
spin-glass problem with equal proportions of ferromagnetic and antiferromagnetic
bonds (Sec.~\ref{sec:clusters_in_the_standard_spin_glass}). Finally,
Sec.~\ref{sec:discussion} contains our conclusions and outlook.

\section{Model, phase diagram and cluster definitions}
\label{sec:model_phase_diagram_and_cluster_definitions}

We study the $\pm J$ random-bond Ising model with Hamiltonian
\begin{align}
\widehat{H}_{\bm{J}}(\bm{S}) = -\sum_{\langle  \bm{x},\bm{y} \rangle } J_{\bm{x} \bm{y} } s_{\bm{x}} s_{\bm{y}}.
\label{eq:hamiltonian}
\end{align}
The Ising spins $s_{ \bm{x} } \in \{ \pm 1 \}$ are placed on the sites of a cubic
lattice of linear size $L$ such that there are $N=L^3$ spins in total. The symbol
$\langle \bm{x}, \bm{y} \rangle$ refers to the summation over nearest neighbors only.
In our actual calculations we always take periodic boundary conditions to remove
surface effects. The exchange couplings $J_{\bm{xy}}$ are time-independent (quenched)
random variables which for the purposes of the present work are drawn from a bimodal
distribution,
\begin{align}
 P_J(J_{\bm{x} \bm{y}} ) = 
   \antifrac \, \delta(J_{\bm{xy}}+1) +
  (1-\antifrac)\, \delta(J_{\bm{xy}}-1).
  \label{eq:bond_distribution}
\end{align}
that is, a bond is either antiferromagnetic $J_{\bm{x} \bm{y}} = -1$ with probability
$\antifrac$ or ferromagnetic $J_{\bm{x} \bm{y}} = 1$ with probability
$1-\antifrac$. The bond is said to be satisfied if
$J_{\bm{x}\bm{y}}s_{\bm{x}}s_{\bm{y}}=1$ and broken if
$J_{\bm{x}\bm{y}}s_{\bm{x}}s_{\bm{y}}=-1$. A key property of this Hamiltonian is the
presence of frustration for any $\antifrac>0$. A system is said to be frustrated if
no spin configuration exists that satisfies all bonds simultaneously.

\begin{figure}
  \begin{center}
    \includegraphics{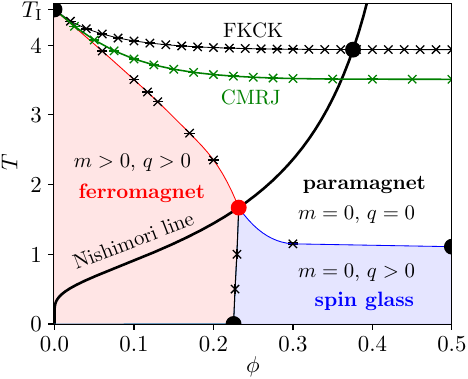}
    \caption{Phase diagram of the three-dimensional $\pm J$ random-bond
      Ising model according to the data of
      Refs.~\cite{HasenbuschEtAl2007CriticalBehaviorOfThe3DPMJIsingModelAtTheParamagneticFerromagneticTransitionLine,HasenbuschEtAl2008CriticalBehaviorOf3DIsingSGModels,HasenbuschEtAl2007MagneticGlassyMulticriticalBehaviorOfThe3DPMJIsingModel,FajenHartmannYoung2020percolationPercolationOfFortuinKasteleynClustersForTheRandomBondIsingModel,CeccarelliPelissettoVivari2011FerromagneticGlassyTransitionsIn3DIsingSGs,CreightonKatzgraber2011SamplingTheGroundStateMagnetizationOfDDimensionalPBodyIsingModels}. The
      data-points for the CMRJ percolation transition originate from the
      present study (green crosses). The red dot illustrates the location
      of the multicritical Nishimori point. The black dots indicate
      important transition points: the pure Ising critical point, the
      intersection of the FKCK transition line with the Nishimori line,
      the spin-glass transition temperature for $\antifrac=0.5$, and the
      zero-temperature transition between the ferromagnetic and
      spin-glass phases. The lines connecting the data points are
      included as visual guides, except for the Nishimori line, which is
      explicitly defined in Eq.~\eqref{eq:nishimori_line}.}
    \label{fig:phase_diagram_pmj_rbim_plot}
  \end{center}
\end{figure}
      
The model is studied in the canonical ensemble at temperature $T$ such that, in
equilibrium, the spin configurations $\bm{S}\in \{ \pm 1 \}^N$ are Gibbs--Boltzmann
distributed. The properties of the thermodynamic state of the model are determined by
two parameters: the temperature $T$ and the fraction of antiferromagnetic bonds
$\antifrac$. Depending on these parameters, the system is in one of four distinct
phases. At high temperatures, the system is in a paramagnetic phase, while at low
temperatures, it transitions either into a ferromagnetic or antiferromagnetic phase,
or into the spin-glass phase, depending on the value of $\antifrac$. Since for the
bipartite simple cubic lattice the antiferromagnetic phase is simply related by
symmetry to the ferromagnetic one, we restrict our attention to the regime
$0\le \phi \le 1/2$.  To differentiate between the paramagnetic, ferromagnetic and
spin-glass phases, two order parameters are introduced: the magnetization, which
characterizes the ferromagnetic state, and the overlap, which describes the
spin-glass state. The magnetization of a single spin configuration is given by
\begin{align}
\widehat{m}( \bm{S})=\frac{1}{N} \sum_{\bm{x}} s_{\bm{x}}.
\end{align}
Within the ferromagnetic phase, the average magnetization
$m:=[ \langle \vert \widehat{m} \vert \rangle_S ]_J $ is non-zero, whereas in the
paramagnetic and spin-glass phases it approaches zero in the thermodynamic limit.
Here, $\langle \ldots \rangle_S$ represents the thermal average taken over the
Gibbs--Boltzmann distribution, while $[\ldots]_J$ denotes the average over the
quenched disorder. The overlap is the order parameter of the spin-glass transition,
and it is defined by
\begin{align} 
 \widehat{q}(\bm{S}^{(1)},\bm{S}^{(2)} ) = \frac{1}{N} \sum_{\bm{x}} q_{\bm{x}} 
\end{align}
where $q_{\bm{x}}=s_{\bm{x}}^{(1)} s_{\bm{x}}^{(2)}$ and
$s_{\bm{x}}^{(1)} \in \bm{S}^{(1)} $, $s_{\bm{x}}^{(2)} \in \bm{S}^{(2)}$. Here
$\bm{S}^{(1)}$ and $\bm{S}^{(2)}$ denote two spin configurations that belong to
different, independent replicas of the system. Accordingly, the average overlap
$q:=[\langle \vert \widehat{q}\, \vert \rangle_{S_2}]_J$ is computed using the
two-replica Gibbs--Boltzmann distribution,
\begin{align}
P_{S_2}(\bm{S}^{(1)},\bm{S}^{(2)}  \vert  \bm{J} ) &=  \frac{1}{Z_{\bm{J}}^2} \exp\!\left( - \widehat{H}_{\bm{J}}^{(2)} /T  \right)~\text{with}\nonumber \\[3pt]
\widehat{H}_{\bm{J}}^{(2)} &= -\sum_{\langle \bm{x}, \bm{y} \rangle} J_{\bm{x}\bm{y}} 
\tilde{s}_{\bm{x}} \tilde{s}_{\bm{y}}.
\label{eq:two_replica_hamiltonian}
\end{align}
Here $\tilde{s}_{\bm{x}}=(s_{\bm{x}}^{(1)},s_{\bm{x}}^{(2)})$ are two-component
vector spins such that
$\tilde{s}_{\bm{x}}\tilde{s}_{\bm{y}}=s_{\bm{x}}^{(1)}s_{\bm{y}}^{(1)}+s_{\bm{x}}^{(2)}s_{\bm{y}}^{(2)}$,
and $Z_{\bm{J}}$ is the partition function for a given realization $\bm{J}$ of bonds. Note that, for convenience, the Boltzmann constant is set to
$k_{\mathrm{B}}:=1$. Asymptotically, the overlap is non-zero in the ferromagnetic and
in the spin-glass phase, whereas it is zero in the paramagnetic phase.

In Fig.~\ref{fig:phase_diagram_pmj_rbim_plot} we show a schematic phase diagram of
the three-dimensional $\pm J$ random-bond Ising
model~\cite{HasenbuschEtAl2009CriticalAndMulticriticalBehaviorOfThePMJIn2And3D},
based on data from
Refs.~\cite{HasenbuschEtAl2007CriticalBehaviorOfThe3DPMJIsingModelAtTheParamagneticFerromagneticTransitionLine,HasenbuschEtAl2007MagneticGlassyMulticriticalBehaviorOfThe3DPMJIsingModel,
  HasenbuschEtAl2008CriticalBehaviorOf3DIsingSGModels,
  FajenHartmannYoung2020percolationPercolationOfFortuinKasteleynClustersForTheRandomBondIsingModel,
  CeccarelliPelissettoVivari2011FerromagneticGlassyTransitionsIn3DIsingSGs,
  CreightonKatzgraber2011SamplingTheGroundStateMagnetizationOfDDimensionalPBodyIsingModels}. The
boundary separating the paramagnetic and ferromagnetic phases starts at the Ising
critical
point~\cite{FerrenbergEtAl2018PushingTheLimitsOfMCSimulationForThe3DIsingModel}
$T_\mathrm{I}:=T_\mathrm{f}(0)=4.511\,523\,26(11)$, which falls within the Ising
universality class. The boundary then extends downward to the multicritical
point~\cite{HasenbuschEtAl2007MagneticGlassyMulticriticalBehaviorOfThe3DPMJIsingModel}
at $T^*=1.6692(3)$ and $\antifrac^{*}=0.231\,80(4)$. In the interval
$0<\antifrac<\antifrac^{*}$, the paramagnetic–ferromagnetic phase transition belongs
to the disordered Ising universality
class~\cite{HasenbuschEtAl2007CriticalBehaviorOfThe3DPMJIsingModelAtTheParamagneticFerromagneticTransitionLine}. At
the multicritical point, also referred to as the Nishimori point, the paramagnetic,
spin-glass, and ferromagnetic phases meet. Beyond this point, where
$\antifrac^{*} < \antifrac \leq 0.5$, the phase transition between the paramagnetic
and spin-glass phases belongs to the Ising spin-glass universality
class~\cite{KatzgraberEtAl2006Universality3DIsingSpinGlasses,HasenbuschEtAl2008CriticalBehaviorOf3DIsingSGModels,BaityJesiEtAl2013CriticalParametersOfThe3DIsingSG}. At
the endpoint of the transition line, where half of the bonds are antiferromagnetic,
the spin-glass transition temperature
is~\cite{BaityJesiEtAl2013CriticalParametersOfThe3DIsingSG}
$T_{\mathrm{sg}}(0.5) = 1.1019(29)$. The boundary between the ferromagnetic and
spin-glass phases extends
from~\cite{CreightonKatzgraber2011SamplingTheGroundStateMagnetizationOfDDimensionalPBodyIsingModels}
$T_{\mathrm{f}}(\antifrac_0)=0$ at $\antifrac_0=0.2253(7)$ up to the multicritical
point~\cite{CeccarelliPelissettoVivari2011FerromagneticGlassyTransitionsIn3DIsingSGs,Hartmann1999GroundStateBehaviorOfThe3DPMJBIsingModel}. The
phase diagram can be symmetrically extended to values of $0.5 < \antifrac \leq 1$. In
this case, the magnetization needs to be replaced by the staggered magnetization in
order to identify antiferromagnetic order instead of ferromagnetic order.

The aim of this study is to characterize thermal phase transitions in the $\pm J$
random-bond Ising model through the properties of several types of clusters. The
corresponding bond-percolation problem is introduced as follows: each bond in the
system is occupied with a probability $p_{\bm{x}\bm{y}}$, and clusters are formed
based on the occupied bonds. Specifically, all spin sites that can be connected
through a path of occupied bonds belong to the same cluster. The smallest possible
cluster consists of just one spin site, while the largest can contain all the
sites. The tuning parameter of percolation is the occupation probability. In the
percolation problems discussed here, the occupation probability depends on the
underlying spin configurations. This contrasts with random percolation, where each
bond is occupied independently with a uniform
probability~\cite{ConiglioFierro2021CorrelatedPercolation}. Typically, there is a
non-percolating subcritical phase for low occupation probabilities and a percolating
supercritical phase for high occupation probabilities. The system is said to
percolate if there are one or more clusters that span the entire system. The
percolation threshold $p_{\mathrm{th}}$ denotes the critical occupation probability
at which an infinite (or incipient percolating) cluster first emerges in the
thermodynamic limit.

A straightforward way to define meaningful clusters in the Ising model is by
identifying regions where the order parameter takes an identical
value~\cite{ConiglioFierro2021CorrelatedPercolation}. For the magnetization, bonds
are hence activated between all like spins, and the bond-occupation probability is
thus given by
\begin{align}
  p_{\bm{x} \bm{y}}^{(\mathrm{Ising})} =  
  \begin{cases}
    1&\text{if}~s_{\bm{x}}=s_{\bm{y}} \\
    0&\text{else}
  \end{cases}
       ,
\end{align}
and we denote the corresponding clusters as Ising clusters. In the same spirit, given
two replicas, geometric clusters can be defined in the domains of constant
\emph{overlap}. In this case, the bond occupation probability is given by
\begin{align}
  p_{\bm{x} \bm{y} }^{(\mathrm{H})} = 
  \begin{cases}
1&\text{if}~q_{\bm{x}}=q_{\bm{y}} \\
0&\text{else}
\end{cases}
   ,
\end{align}
and we denote the corresponding clusters as Houdayer
clusters~\cite{Houdayer2001ClusterMonteCarloAlgorithmFor2DimensionalSpinGlasses}. Although
both Ising and  Houdayer clusters are useful to describe the phases of the
$\pm J$ random-bond Ising model, they do not properly represent the correlations of
the order parameters. Clusters that have this property, at least when the system is
not frustrated, are the single-replica FKCK clusters with bond-occupation
probability~\cite{FortuinKasteleyn1972OnTheRandomClusterModel1,ConiglioKlein1980ClustersAndCriticalDroplets}
\begin{align}
p_{\bm{x}\bm{y}}^{(\mathrm{FKCK})} =
\begin{cases}
1-\exp\!\left(-\dfrac{2J_{\bm{x}\bm{y}} s_{\bm{x}} s_{\bm{y}}}{T}\right) & \text{if } J_{\bm{x}\bm{y}} s_{\bm{x}} s_{\bm{y}} > 0 \\[3pt]
0 & \text{else}
\end{cases}
.
\label{eq:occupation_probability_fkck_clusters}
\end{align}
For the pure Ising model at $\antifrac=0$, the FKCK percolation transition maps onto
the ferromagnetic phase transition, see Eq.~\eqref{eq:Ising-FKCK-correspondence},
exhibiting identical critical
behavior~\cite{ConiglioFierro2021CorrelatedPercolation}. However, for $\antifrac>0$,
this mapping breaks down, cf.~Eq.~\eqref{eq:broken-Ising-FKCK-correspondence}, and
the critical temperature of the percolation transition exceeds that of both the
ferromagnetic and spin-glass
transitions~\cite{ConiglioEtAl1991ClusterApproachToSpinGlassesAndTheFrustratedPercolationProblem,CataudellaEtAl1994CriticalClustersAndEfficientDynamicsForFrustratedSpinModels,DeSantisGandolfi1999BondPercolationInFrustratedSystems}. According
to Monte Carlo simulations, the FKCK percolation transition for values of
$\antifrac>0$ belongs to the universality class of random
percolation~\cite{DeArcangelis1991PercolationTransitionInSpinGlasses,FajenHartmannYoung2020percolationPercolationOfFortuinKasteleynClustersForTheRandomBondIsingModel}. On
the Nishimori line, the critical temperature of the FKCK percolation transition can
be determined using the random-bond percolation threshold of the cubic lattice, given
by~\cite{WangEtAl2013BondAndSitePercolationInThreeDimensions}
$p_{\mathrm{th}}=0.248\,811\,82(10)$. The Nishimori line is a special set in
parameter space derived via a gauge transformation. On this line, physical quantities
such as the internal energy can be computed exactly and remarkable correlation
identities
hold~\cite{Nishimori1980ExactResultsAndCriticalPropertiesOfTheIsingModelWithCompetingInteractions,Nishimori1981InternalEnergySpecificHeatAndCorrelationFunctionOftheBondRandomIsingModel}. The
line originates at $T=0$ for $\antifrac=0$, passes through the multicritical point,
and diverges as $ \antifrac \to 0.5$. Its precise form is defined by
\cite{Nishimori1981InternalEnergySpecificHeatAndCorrelationFunctionOftheBondRandomIsingModel}
\begin{align}
T_\mathrm{N}(\antifrac)=\frac{2}{\ln \! \left[ (1-\antifrac)/\antifrac \right]}.
\label{eq:nishimori_line}
\end{align} 
At $\antifrac_{\mathrm{N/FKCK}}=(1-p_{\mathrm{th}})/2$ the Nishimori line intersects
the FKCK transition
line~\cite{Yamaguchi2013ConjecturedExactPercolationThresholdsOfTheFortuinKasteleynClusters,Yamaguchi2010PercolationThresholdsOfTheFortuinKasteleynClusterForTheEAIsingModelOnComplexNetworks,Gandolfi1999ARemarkOnGaugeSymmetriesInIsingSpinGlasses}
as shown in Fig.~\ref{fig:phase_diagram_pmj_rbim_plot}. At this intersection, it has
been rigorously proven that the FKCK transition belongs to the random-percolation
universality class~\cite{Gandolfi1999ARemarkOnGaugeSymmetriesInIsingSpinGlasses}.

For $\antifrac > 0$, the physical interpretation of the FKCK percolation transition
remains
unclear~\cite{Gandolfi1999ARemarkOnGaugeSymmetriesInIsingSpinGlasses,NewmanStein2007ShortRangeSpinGlassesResultsAndSpeculations,LundowCampbell2012FKAndDamageSpreadingTransitionsInRandomBondIsingLattices}. Instead,
a more meaningful approach is to consider clusters defined across multiple replicas
that can be directly linked to the
overlap~\cite{SwendsenWang1986ReplicaMonteCarloSimulationOfSpinGlasses}, such as 
Houdayer clusters. A two-replica cluster definition, in this spirit, emerges from the
graphical representation of Chayes, Machta, and
Redner~\cite{CMR1998GraphicalRepresentationsForIsingSystemsInExternalFields,MachtaNewmanStein2008ThePercolationSignatureOfTheSpinGlassTransition,MachtaNewmanStein2009APercolationTheoreticApproachToSGPhaseTransitions},
as well as from the cluster algorithm of
J\"org~\cite{Joerg2005ClusterMonteCarloAlgorithmsForDilutedSpinGlasses}. The
bond-occupation probability of these CMRJ clusters is given by~\footnote{In the
  graphical representation of Chayes, Machta and Redner the CMRJ clusters are
  actually denoted as blue clusters. Here, we use the acronym CMRJ for blue
  clusters~\cite{MachtaNewmanStein2008ThePercolationSignatureOfTheSpinGlassTransition}.}
\begin{align}
p_{\bm{x}\bm{y}}^{(\mathrm{CMRJ})} =
\begin{cases}
1-\exp\!\left(-\dfrac{2J_{\bm{x}\bm{y}} \tilde{s}_{\bm{x}} \tilde{s}_{\bm{y}}}{T}\right) & \text{if } J_{\bm{x}\bm{y}} \tilde{s}_{\bm{x}} \tilde{s}_{\bm{y}} > 0 \\[3pt]
0 & \text{else}
\end{cases}
.
\label{eq:occupation_probability_cmrj_clusters}
\end{align}
This implies that a bond is occupied only if it is simultaneously satisfied in both
replicas. Consequently, spins within the same CMRJ cluster share an identical value
of the overlap. The results for the percolation temperatures of the CMRJ clusters,
presented in Fig.~\ref{fig:phase_diagram_pmj_rbim_plot}, indicate that these clusters
percolate within the temperature range between the FKCK percolation transition and
the thermal phase transition for $0<\antifrac \leq 0.5$. In the special case of
$\antifrac = 0$, CMRJ clusters percolate precisely at the critical temperature of the
pure ferromagnet, $T_{\mathrm{I}}$.

Note that Newman and Stein also proposed a definition for multi-replica clusters in
spin glasses, the two-replica FKCK clusters~\cite{NewmanStein2007ShortRangeSpinGlassesResultsAndSpeculations}. These
ideas are further discussed in Appendix~\ref{sec:nreplica_fkck_percolation}.

\section{Observables and finite size scaling}
\label{sec:observables_and_finite_size_scaling}

In order to investigate the connection between the thermal (ferromagnetic or
spin-glass) phase transitions in the disordered Ising model and the percolation
transitions of the various types of clusters considered, we study observables that
are able to describe magnetic and spin-glass ordering as well as quantities typically
considered in percolation theory.

\subsection{Percolation quantities}

Clusters naturally are the central objects of interest in
percolation~\cite{StaufferAharony1994IntroductionToPercolationTheory,SahimiHunt2021BookComplexMediaAndPercolationTheory,Grimmett1999Percolation}. The
density $\widehat{\rho}$ of a cluster is defined as the number of sites contained in
the cluster, also referred to as the cluster size, divided by the total number $N$ of
lattice sites in the system. In the following, we will assume that the clusters are
ordered by their densities such that $\widehat{\rho}_i \geq \widehat{\rho}_{i+1}$ for
$i=1,2,3,\ldots$, meaning that $\widehat{\rho}_1$ represents the largest cluster,
$\widehat{\rho}_2$ the second largest, and so forth. The order parameter of the
percolation transition is the density of the largest cluster,
$\rho_1 = [ \langle \widehat{\rho}_1 \rangle_{\mathrm{cl}} ]_J$, also known as the
percolation strength $P_\infty$, where $\langle \ldots \rangle_{\mathrm{cl}}$ denotes
the average over the configurations of occupied
bonds~\cite{StaufferAharony1994IntroductionToPercolationTheory}.

The tuning parameter of the percolation transition is the occupation probability. For
the clusters defined in
Section~\ref{sec:model_phase_diagram_and_cluster_definitions}, the occupation
probabilities explicitly or implicitly depend on temperature. As a result, the
observables are studied as functions of temperature. According to finite-size
scaling, the leading-order singular behavior of the largest cluster in the critical
region is given by~\cite{StaufferAharony1994IntroductionToPercolationTheory}
\begin{align}
 L^{\beta/\nu}  \rho_1 = \Psi_{\rho} \! \left(  t L^{1/ \nu} \right),
\label{eq:fss_largest_cluster_size}
\end{align}
where $t=(T-T_{\mathrm{c}})/T_{\mathrm{c}}$ is the reduced temperature and
$\Psi_{\rho}$ is the corresponding scaling function. Here $T_{\mathrm{c}}$ denotes
the critical temperature of the percolation transition of the considered clusters,
which is analogous to the percolation threshold in random percolation. At the
critical temperature, the average cluster size diverges
as~\cite{StaufferAharony1994IntroductionToPercolationTheory}
\begin{align}
\chi_{\rho} ( T_{\mathrm{c}} ) \sim L^{\gamma/\nu },
\label{eq:power_law_average_cluster_size}
\end{align}
where
\begin{align*}
\chi_{\rho}= N \left[ \, \left \langle \sum_{i=1} \widehat{\rho}_i^{\,2} \right \rangle_{\mathrm{cl}} \,\right]_J.
\end{align*}
The sum runs over the densities of all clusters. The percolation transition separates
the high-temperature phase, characterized by local clusters only, from the
percolating phase, where there is at least one system-spanning cluster. In this work,
only finite-size systems with periodic boundary conditions in all directions are
considered, and the criterion for percolation is satisfied if a cluster wraps around
the boundaries in at least one direction. The wrapping probability, thus, provides
information about the location of the percolation
transition~\cite{NewmanZiff2001FastMonteCarloAlgorithmForSiteOrBondPercolation}. The
wrapping probability is a dimensionless quantity that follows the scaling
form~\cite{StaufferAharony1994IntroductionToPercolationTheory}
\begin{align}
R(t,L) = \Psi_R \! \left( t L^{1/\nu} \right).
\label{eq:fss_wrapping_probability}
\end{align}
If there is more than one wrapping cluster, the number of wrapping clusters $w_{R}$
can also be studied as an observable. The number of wrapping clusters obeys the same
scaling law as the wrapping probability.

\subsection{Thermal quantities}

To establish a connection between geometric and thermal properties, it is necessary
to investigate the behavior of the thermal order parameters $m$ and $q$. To leading
order, $m$ and $q$ satisfy the scaling
laws~\cite{Janke2008MonteCarloMethodsInClassicalStatisticalPhysics}
\begin{equation}
 L^{\beta/\nu}  m = \Psi_{m} \! \left(  t L^{1/ \nu} \right), 
  \quad
   L^{\beta/\nu}  q = \Psi_{q} \! \left(  t L^{1/ \nu} \right),
\label{eq:fss_thermal_order_parameter}
\end{equation}
in the critical region of the respective phase transition. Specifically, this
corresponds to a ferromagnetic transition in the case of the magnetization and to a
spin-glass transition in the case of the overlap. At the critical point, the
susceptibility diverges
as~\cite{Janke2008MonteCarloMethodsInClassicalStatisticalPhysics}
\begin{align}
\chi_{m/q}(T_{\mathrm{c}}) \sim L^{\gamma/\nu},  
\label{eq:power_law_susceptibility}
\end{align}
where $\chi_m = N \bigl[ \langle \widehat{m}^{2} \rangle_S \bigr]_J$ or
$\chi_q = N \bigl[ \langle \widehat{q}^{\,2} \rangle_{S_2} \bigr]_J$, depending on
the transition. Furthermore, in the critical region, the correlation length
$\xi_{m/q}$ behaves as~\cite{NishimoriOrtiz2011RenormalizationGroupAndScaling}
\begin{align}
\frac{\xi_{m/q}(t,L)}{L} = \Psi_{\xi_{m/q}} \! \left( t L^{1/\nu} \right).  
\label{eq:fss_correlation_length}
\end{align} 
The correlation length is computed using the second-moment estimator, which is based
on the Fourier transform $\widetilde{g}_{m/q}(\bm{k})$ of the two-point correlation
function $g_{m/q}(\bm{r})$,
i.e., \cite{Janke2008MonteCarloMethodsInClassicalStatisticalPhysics}
\begin{align}
\xi_{m/q} = \frac{1}{2\sin \!{( k_{\mathrm{min}} /2 )}} \sqrt{\frac{\widetilde{g}_{m/q}(\bm{0})- \widetilde{g}_{m/q}(\bm{k}) }{ \widetilde{g}_{m/q}(\bm{k})}}, 
\label{eq:second_moment_correlation_length}
\end{align}
where $\bm{k}=(k_{\mathrm{min}},0,0)$, $k_{\mathrm{min}}=2\pi/L$, and
$\bm{0}=(0,0,0)$. Depending on the context, the correlations of the magnetization,
$g_{m}(\bm{r}) = [\langle s_{\bm{0}} s_{\bm{0}+\bm{r}} \rangle_S]_J$, or of the
overlap, $g_{q}(\bm{r}) = [\langle q_{\bm{0}} q_{\bm{0}+\bm{r}} \rangle_{S_2}]_J$,
are considered.

\subsection{Simulations and finite-size scaling analysis}

The thermal and geometrical properties of the $\pm J$ random-bond Ising model as
characterized by the above sets of corresponding observables are studied utilizing
parallel-tempering Monte Carlo simulations. To achieve good equilibration, we employ
a combination of single-spin flip, replica-exchange and cluster-update moves. All
data are averaged over a large number of disorder realizations in order to bring down
statistical errors to an acceptable level. More information regarding the simulation
scheme, including details about the system sizes considered and the number of
disorder samples employed, are provided in Appendix~\ref{sec:simulation_details}.

The numerical strategy for extracting the critical temperature and critical exponents
is as follows. In the first step, a data collapse of the wrapping probability or the
number of wrapping clusters is carried out in order to determine $T_{\mathrm{c}}$ and
$\nu$ of a percolation transition, according to
Eqs.~\eqref{eq:fss_wrapping_probability}. To obtain $T_{\mathrm{c}}$ and $\nu$ of a
thermal phase transition, a similar data collapse is performed using the correlation
length and Eq.~\eqref{eq:fss_correlation_length}. In the second step, a power law is
fitted to the data of the average cluster size or the susceptibility at the
previously estimated value of the critical temperature $T_{\mathrm{c}}$, yielding
$\gamma/\nu$ via Eq.~\eqref{eq:power_law_average_cluster_size} or
Eq.~\eqref{eq:power_law_susceptibility}, respectively. Since a data point at
$T_{\mathrm{c}}$ is not necessarily available, a linear interpolation of the relevant
quantity is carried out to obtain the value. To extract the critical exponent
$\beta/\nu$, a data collapse of the order parameter is performed according to the
scaling law in Eq.~\eqref{eq:fss_largest_cluster_size} or
Eq.~\eqref{eq:fss_thermal_order_parameter}, fixing $1/\nu$ to the previously
determined value. The optimal parameters for the data collapses are extracted using
the tool provided in Ref.~\cite{Melchert2009Autoscale}. In order to calculate the
statistical errors of the adjustable parameters of the data collapses, bootstrapping
is applied~\cite{Young2015EverythingYouWantedToKnowAboutDataAnalysis}. To estimate
the error from corrections to scaling, the minimal system size included in the
collapses is altered, and the argument range of the collapse is varied within the
interval
$|t| L^{1/\nu} \lesssim 1$~\cite{NishimoriOrtiz2011RenormalizationGroupAndScaling}.

Our analysis focuses on the behavior of the $\pm J$ random-bond Ising model for three
distinct fractions of antiferromagnetic bonds, $\antifrac = 0$, $\antifrac = 0.125$,
and $\antifrac = 0.5$. We first focus on the reference case $\antifrac = 0$.

\begin{figure}
	\begin{center}	
           \includegraphics{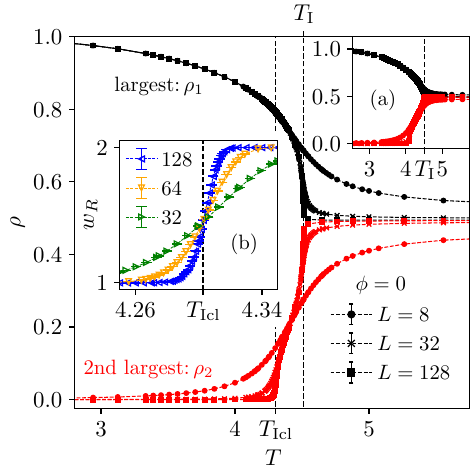}
	\end{center}
        \caption{Densities of the two largest Ising clusters as a function of
          temperature for the pure ferromagnet using three exemplary system sizes,
          $L=8$, $L=32$ and $L=128$. Inset (a) shows the corresponding relationship
          for the Houdayer clusters. $T_{\mathrm{I}}$ denotes the critical
          temperature of the ferromagnetic phase transition. Inset (b) illustrates
          how the number of wrapping Ising clusters decreases from two to one at
          $T_{\mathrm{Icl}}$ (see main text for details). The system sizes plotted
          are $L = 32$, $64$, and $128$.}
        \label{fig:p0_ising_clusters_and_houdayer_clusters_plot}
\end{figure}

\section{Clusters in the pure ferromagnet} 
\label{sec:clusters_in_the_pure_ferromagnet}

\begin{figure}[tb!]
	\begin{center}
		\includegraphics{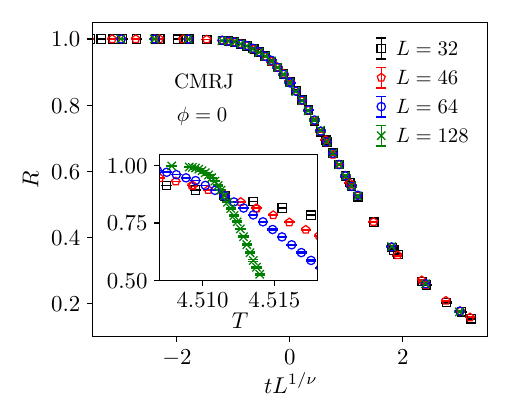}
	\end{center}
        \caption{Data collapse of the wrapping probability of CMRJ clusters, obtained
          according to Eq.~\eqref{eq:fss_wrapping_probability}, in the critical
          region of the CMRJ percolation transition for the pure Ising ferromagnet
          ($\antifrac = 0$), with $t=(T-T_\mathrm{CMRJ})/T_\mathrm{CMRJ}$. From this
          collapse, the critical parameters are estimated as
          $T_\mathrm{CMRJ}=4.511\,527(16)$ and $\nu=0.6300(17)$. The inset shows the
          unscaled data in the vicinity of the critical point.}
        \label{fig:p0_cmrj_wrapping_prob_plot}
\end{figure}

When $\antifrac=0$, there are no antiferromagnetic bonds present, and the Hamiltonian
in Eq.~\eqref{eq:hamiltonian} reduces to that of a pure, unfrustrated Ising
ferromagnet. This model undergoes a phase transition from the paramagnetic to the
ferromagnetic phase
at~\cite{FerrenbergEtAl2018PushingTheLimitsOfMCSimulationForThe3DIsingModel}
$T_{\mathrm{I}}=4.511\,523\,26(11)$. As illustrated in
Fig.~\ref{fig:p0_ising_clusters_and_houdayer_clusters_plot}, this phase transition is
visible in both the Ising and Houdayer
clusters~\cite{Mueller1974PercolationInALatticeSystemWithParticleInteraction}. The
figure shows the behavior of the largest and second largest cluster as a function of
temperature. At high temperatures, there are two giant percolating clusters with
equal densities. The reason is that for $T \to \infty$, spin orientations are random,
so on average half of the spins are up and the other half down, causing both Houdayer
and Ising clusters to behave similarly to a random-site percolation problem. The
site-percolation threshold on a cubic lattice
is~\cite{WangEtAl2013BondAndSitePercolationInThreeDimensions}
$p_{\mathrm{th}}^{(\mathrm{s})}=0.311\,607\,7(2) < 1/2$. As a result, there are two
percolating clusters, one with positive and the other with negative values of the
magnetization or overlap, respectively. This argument remains valid for any value of
$\antifrac$ when $T\to \infty$. At the ferromagnetic phase transition, one cluster
begins to dominate the other, and the densities start to deviate from
$\rho_{1/2} \approx 0.5$. Inset (b) of
Fig.~\ref{fig:p0_ising_clusters_and_houdayer_clusters_plot} shows how the number of
wrapping Ising clusters decreases from two to one at the temperature~\footnote{The
  percolation temperature $T_{\mathrm{Icl}} = 4.3027(3)$, as well as the critical
  exponent $\nu = 0.882(5)$, are extracted by performing a data collapse of the
  number of wrapping Ising clusters, i.e., $w_R(t,L) = \Psi_w(t L^{1/\nu})$, where
  $t = (T - T_{\mathrm{Icl}})/T_{\mathrm{Icl}}$ (plot not shown). The exponent is
  consistent with random percolation.}~$T_{\mathrm{Icl}}=4.3027(3)$. Notably, this
temperature is lower than the critical temperature of the ferromagnetic phase
transition~\cite{Mueller1974PercolationInALatticeSystemWithParticleInteraction}. Hence,
Ising clusters clearly do not encapsulate the critical behavior of the thermal phase
transition. In contrast, in two dimensions the second largest cluster percolates
right at $T_\mathrm{I}$, but the corresponding critical exponents nevertheless differ
from those of the thermal phase transition~\cite{stella:89}.  At zero temperature,
the magnetization and the overlap reach unity, implying that there is a single Ising
or Houdayer cluster containing all spin sites.

Although both Ising and Houdayer clusters reflect the ordering behavior of the model,
their percolation properties do not map onto the ferromagnetic phase
transition~\cite{Mueller1974PercolationInALatticeSystemWithParticleInteraction,Fortunato2003ClusterPercolationAndCriticalBehaviourInSpinModels,DeLaRochaEtAl2015DomainSizeHeterogeneityInTheIsingModel}. Clusters
that do exhibit this property, at least for $\antifrac=0$, are the FKCK clusters. For
them, the spin--spin correlation function is equivalent to the connectivity function
of the corresponding percolation problem, cf.\
Eq.~\eqref{eq:Ising-FKCK-correspondence}~\cite{ConiglioFierro2021CorrelatedPercolation,Grimmett2004TheRandomClusterModel}.
In other words, the correlation between spins $s_{\bm{x}}$ and $ s_{\bm{y}}$ is
identical to the probability that the lattice sites $\bm{x}$ and $\bm{y}$ belong to
the same FKCK cluster. In general, the behavior of the FKCK clusters differs
significantly from that of the previously discussed Ising and Houdayer clusters. At
high temperatures, percolation does not occur because the occupation probability
depends on temperature and vanishes for $T\to \infty$. Due to
Eq.~\eqref{eq:Ising-FKCK-correspondence}, the critical behavior of the FKCK
percolation transition is identical to that of the ferromagnetic phase
transition. Below the transition temperature, there is a unique largest cluster whose
density equals the magnetization in the thermodynamic
limit~\cite{ConiglioFierro2021CorrelatedPercolation}.

Interestingly, we observe that for the case of $\antifrac = 0$ the two-replica CMRJ
clusters exhibit the same percolation behavior as the FKCK
clusters. Figure~\ref{fig:p0_cmrj_wrapping_prob_plot} illustrates the finite-size
scaling behavior of the wrapping probability of CMRJ clusters. From the data
collapse, the critical temperature $T_{\mathrm{CMRJ}}=4.511\,527(16)$ and the
critical exponent $\nu=0.6300(17)$ are obtained.

\begin{figure}[tb!]
	\begin{center}
		\includegraphics{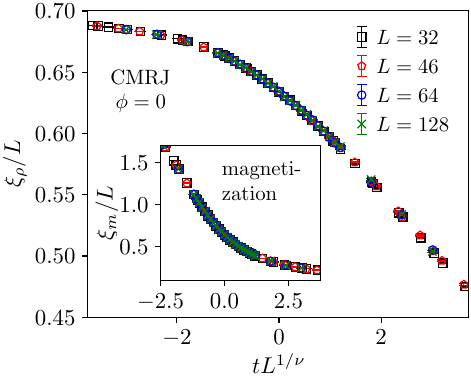}
	\end{center}
        \caption{Data collapse of the connectivity length according to
          Eq.~\eqref{eq:cluster_connectivity_length} with
          $t=(T-T_\mathrm{CMRJ})/T_{\mathrm{CMRJ}}$ for the parameters
          $T_\mathrm{CMRJ}=4.511\,527$ and $\nu=0.6300$. The inset shows the
          corresponding data collapse for the correlation length of the
          magnetization, defined in Eq.~\eqref{eq:second_moment_correlation_length},
          using the same parameters, i.e., $T_\mathrm{I}=4.511\,527$, $\nu=0.6300$,
          and $t=(T-T_\mathrm{I})/T_\mathrm{I}$. }
        \label{fig:p0_cmrj_and_magnetization_correlation_length_fss_plot}
\end{figure}

To determine the exponent $\gamma/\nu$, power-law fits are performed using
Eq.~\eqref{eq:power_law_average_cluster_size} for the average cluster size at the
temperature~\cite{FerrenbergEtAl2018PushingTheLimitsOfMCSimulationForThe3DIsingModel}
$T=4.511\,523\,256$. For small system sizes, we notice the presence of scaling
corrections for this quantity, and it is necessary to take these into account in the
fits. In the present context, we achieve this by performing a power-extrapolation of
effective critical exponents (local slopes) found from fits on a restricted range of
system sizes. The detailed procedure is described in
Appendix~\ref{sec:extrapolation_ansatz_for_scaling_corrections}, where we also
include a plot showing the scaling of the average cluster size at the critical
temperature. The resulting estimate for the average cluster-size (susceptibility)
exponent for the CMRJ clusters is $\gamma/\nu=1.9638(18)$.

To directly test whether the correlations of the ferromagnetic phase transition and
the CMRJ percolation transition behave identically in the vicinity of the phase
transition, the critical behavior of the correlation length is compared to that of
the cluster connectivity length. The connectivity length can be computed
as~\cite{StaufferAharony1994IntroductionToPercolationTheory}
\begin{align}
   \xi_{\rho}= \sqrt{
   \frac{2N \left [ \left \langle \sum_{i=1} \widehat{r}_i^{\,2} \widehat{\rho}_i^{\,2} \right \rangle_{\mathrm{cl}} \right ]_J }{
   \chi_{\rho}  
   } }.
\end{align}
Here, $\widehat{r}_i$ denotes the radius of the $i$th largest cluster. The radius of
a cluster containing $n$ lattice sites is given
by~\cite{StaufferAharony1994IntroductionToPercolationTheory}
\begin{align}
 \widehat{r}=\sqrt{ \frac{1}{n} \sum_{j=1}^{n}  {\vert \bm{r}_j - \bm{r}_{\mathrm{0}} \vert}^{2} }.
\end{align}
The summation is performed over the Euclidean distances between the positions of the
lattice sites in the cluster $\bm{r}_j$ and the center of mass of the cluster
$\bm{r}_{\mathrm{0}}$. For non-wrapping clusters, the center of mass can be defined
with respect to any lattice position. For wrapping clusters, an algorithm proposed in 
Ref.~\cite{BaiBreen2008CalculatingCenterOfMassInAnUnbounded2DEnvironment} is used to
determine the center of mass. Analogous to the correlation length, the connectivity
length follows the scaling
form~\cite{StaufferAharony1994IntroductionToPercolationTheory}
\begin{align}
   \frac{\xi_{\rho}(t,L)}{L} = f_{\xi_{\rho}} \!  \left( t L^{1/\nu} \right).
   \label{eq:cluster_connectivity_length}
\end{align}
Figure~\ref{fig:p0_cmrj_and_magnetization_correlation_length_fss_plot} shows the data
collapses of the connectivity length of the CMRJ clusters and the correlation length
of the magnetization, using Eq.~\eqref{eq:fss_correlation_length} and
Eq.~\eqref{eq:cluster_connectivity_length}, respectively. The collapse parameters
were previously obtained from the wrapping probability of the CMRJ clusters, see
Fig.~\ref{fig:p0_cmrj_wrapping_prob_plot}. Figure~\ref{fig:p0_cmrj_and_magnetization_correlation_length_fss_plot}
therefore directly demonstrates that the connectivity length and the correlation
length exhibit identical critical behavior.

\begin{figure}[tb!]
	\begin{center}
		\includegraphics{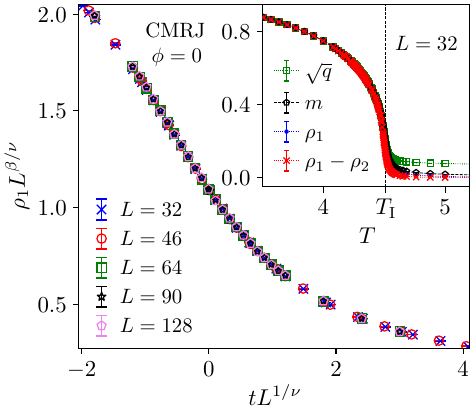}
     \end{center}     
     \caption{Data collapse of the density of the largest CMRJ cluster according to
       Eq.~\eqref{eq:fss_largest_cluster_size} for the pure Ising ferromagnet with
       $t=(T-T_\mathrm{CMRJ})/T_\mathrm{CMRJ}$. From this collapse, the values
       $T_{\mathrm{CMRJ}} = 4.511\,534(26)$ and $\beta/\nu = 0.516(4)$ are
       obtained. The parameter $1/\nu$, with $\nu = 0.6300$, is held constant during
       the optimization. The inset illustrates four different quantities, all of
       which can serve as order parameters of the ferromagnetic phase transition.}
    \label{fig:p0_magnetization_and_cmrj_largest_clusters_plot}
\end{figure}

As mentioned above, the density of the largest FKCK cluster equals the magnetization
in the thermodynamic
limit~\cite{ConiglioFierro2021CorrelatedPercolation,WeigelEtAl2002RandomClusterMultihistogramSamplingForTheQStatePottsModel}. In
other words,
\begin{align}
\rho_1 = m,
\end{align}
for $L \to \infty$. This property also holds for the CMRJ clusters, as illustrated in
Fig.~\ref{fig:p0_magnetization_and_cmrj_largest_clusters_plot}. The inset of this
figure shows four different quantities that exhibit the same behavior as functions of
temperature and can serve as order parameters of the ferromagnetic phase
transition. Notably, the density of the second-largest CMRJ cluster vanishes for
$L \to \infty$, so that
$\rho_1-\rho_2=\rho_1$. Figure~\ref{fig:p0_magnetization_and_cmrj_largest_clusters_plot}
also shows that, up to corrections,
\[
  m = \sqrt{q}
\]
in the ordered phase. This identity follows from taking a pure-state average of a
spin
$s_\mathbf{x}$ for the case of translational invariant systems (recall that we are
using periodic boundary conditions). In the high-temperature limit
$T\to\infty$, in contrast, one sees that $m = q =
1/\sqrt{N}$.  From the data collapses of the largest CMRJ cluster shown in the main
plot of Fig.~\ref{fig:p0_magnetization_and_cmrj_largest_clusters_plot}, the critical
exponent $\beta/\nu =
0.516(4)$ is estimated. Note that in this case only the critical temperature and the
exponent $\beta/\nu$ are optimized, while the exponent
$1/\nu$ is fixed to the previously determined value. This choice is made because the
optimization method is more stable if only two parameters are
varied~\cite{Melchert2009Autoscale}.

\begin{table}[tb!]
  \caption{Critical quantities of the pure Ising ferromagnet obtained from the
    correlation length and the susceptibility of the magnetization, as well as the
    average cluster size and the wrapping probability of the FKCK and CMRJ clusters,
    respectively. To extract $\gamma/\nu$ we have performed power-law fits at
    $T=4.511\,523\,256$ to the data of the susceptibility and the average cluster
    size according to Eq.~\eqref{eq:power_law_susceptibility} and
    Eq.~\eqref{eq:power_law_average_cluster_size}, respectively. The results are in
    agreement with the more accurate values given in
    Ref.~\cite{FerrenbergEtAl2018PushingTheLimitsOfMCSimulationForThe3DIsingModel},
    i.e., $T_{\mathrm{c}}=4.511\,523\,26(11)$, $\nu=0.629\,912(86)$ and
    $\gamma/\nu =1.963\,90(45)$. }
  \begin{tabular}{ c  c  c  c }
    \toprule
    & magnetization & FKCK & CMRJ \\ \hline
    $T_\mathrm{c}$&4.511\,537(12) & 4.511\,522(19)   & 4.511\,527(16) \\
    $\nu$& 0.6310(23) & 0.630(4)    & 0.6300(17)\\
    $\gamma/\nu$& 1.965(4)  & 1.9630(21)  & 1.9638(18) \\ \botrule
  \end{tabular}
  \label{tab:critical_exponents_p0}
\end{table}

In summary, the estimates collected in Table~\ref{tab:critical_exponents_p0} indicate that the CMRJ percolation transition belongs to the three-dimensional Ising universality class. This suggests that in the case of a pure Ising ferromagnet, the CMRJ clusters exhibit the same percolation properties as the FKCK clusters. The reason for this is that the correlation function of the two-replica ferromagnet with Hamiltonian
\begin{align}
\widehat{H}_{\mathrm{I}}^{(2)} = - \sum_{\langle \bm{x}, \bm{y} \rangle }  \tilde{s}_{\bm{x}} \tilde{s}_{\bm{y}}
\label{eq:two_replica_ferromagnet}
\end{align}
is proportional to that of a single ferromagnet at the same temperature~\cite{Suzuki1967SolutionOfPottsModelForPhaseTransition,AizenmanSimon1980AComparisonOfPlaneRotorAndIsingModels}. Furthermore, the CMRJ clusters can be intuitively interpreted as a two-replica generalization of the FKCK clusters, adapted to the Hamiltonian $\widehat{H}_{\mathrm{I}}^{(2)}$, see Eqs.~\eqref{eq:occupation_probability_fkck_clusters} and \eqref{eq:occupation_probability_cmrj_clusters}. As a result, the CMRJ percolation transition maps onto the ferromagnetic phase transition. A more detailed derivation of this result can be found in Appendix~\ref{sec:correlations_and_connectivity_in_cmrj_clusters}.


\section{Clusters in the frustrated disordered ferromagnet}
\label{sec:clusters_in_the_frustrated_disordered_ferromagnet}

The transition between the paramagnetic and ferromagnetic phases for
$0 < \antifrac < \antifrac^{*}$ falls within the universality class of the disordered
Ising ferromagnet,
where~\cite{HasenbuschEtAl2007MagneticGlassyMulticriticalBehaviorOfThe3DPMJIsingModel}
$\antifrac^{*}=0.231\,80(4)$ denotes the location of the multicritical
point~\cite{HasenbuschEtAl2007CriticalBehaviorOfThe3DPMJIsingModelAtTheParamagneticFerromagneticTransitionLine}. Due
to frustration, the correspondence between the connectivity of FKCK clusters and the
spin--spin correlations given in Eq.~\eqref{eq:Ising-FKCK-correspondence} is no
longer
valid~\cite{CataudellaEtAl1994CriticalClustersAndEfficientDynamicsForFrustratedSpinModels}.
As a consequence, for such systems the connectivity length $\xi_{\rho}$ is larger
than the correlation length
$\xi_{m/q}$~\cite{CataudellaEtAl1994CriticalClustersAndEfficientDynamicsForFrustratedSpinModels}.

\begin{figure}[tb!]
	\begin{center}
		\includegraphics{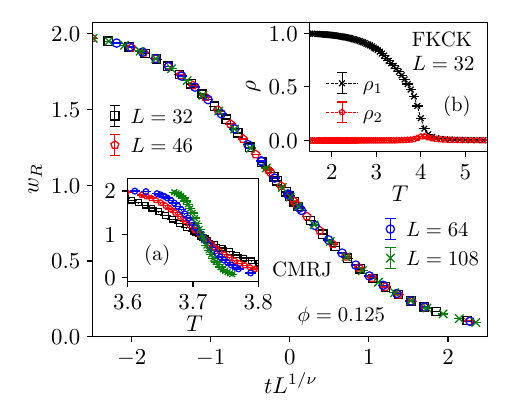}
	\end{center}
        \caption{The main plot shows the number of wrapping CMRJ clusters for
          $\antifrac=0.125$, rescaled according to the scaling law of
          Eq.~\eqref{eq:fss_wrapping_probability} with
          $t=(T-T_\mathrm{CMRJ})/T_{\mathrm{CMRJ}}$. The critical quantities are
          estimated as $T_{\mathrm{CRMJ}}=3.71523(24)$ and $\nu=0.875(9)$. Inset (a)
          illustrates the unscaled data. Inset (b) shows the two largest FKCK
          clusters for $\antifrac=0.125$ at system size $L=32$. Notably there is only
          one wrapping FKCK cluster.}
        \label{fig:p0125_cmrj_percolation_plot}
\end{figure}

We performed extensive simulations of the system with $\antifrac=0.125$. Following
the finite-size scaling protocol outlined in
Sec.~\ref{sec:observables_and_finite_size_scaling}, we extract an estimate
$T_{\mathrm{f}}=3.2412(4)$ for the ferromagnetic ordering transition. In contrast,
the FKCK percolation transition occurs at $T_{\mathrm{FKCK}}=4.02079(10)$. This
transition exhibits the same characteristics as a random percolation transition, with
a single infinite cluster in the percolating phase, as shown in the inset (b) of
Fig.~\ref{fig:p0125_cmrj_percolation_plot}. The critical exponents of the transition
fall within the random percolation universality
class~\cite{DeArcangelis1991PercolationTransitionInSpinGlasses,FajenHartmannYoung2020percolationPercolationOfFortuinKasteleynClustersForTheRandomBondIsingModel},
cf.\ the data collected in Table~\ref{tab:critical_exponents_p0125}.

As shown in the phase diagram in Fig.~\ref{fig:phase_diagram_pmj_rbim_plot}, the CMRJ
percolation transition occurs in the temperature range between the FKCK percolation
point and the ferromagnetic phase transition.  Unlike the FKCK transition, the CMRJ
transition involves the formation of two wrapping clusters, as demonstrated by the
data collapse in Fig.~\ref{fig:p0125_cmrj_percolation_plot}.  The critical
temperature is $T_{\mathrm{CMRJ}}=3.71523(24)$.  The CMRJ percolation transition also
belongs to the random percolation universality class.
Table~\ref{tab:critical_exponents_p0125} summarizes the critical quantities of the
three transitions.

\begin{figure}[tb!]
	\begin{center}
		\includegraphics{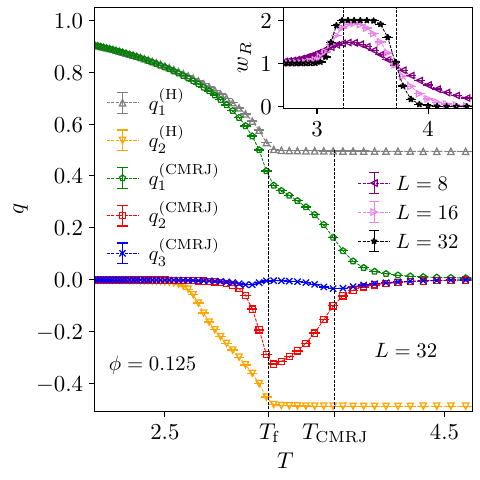}
	\end{center}
        \caption{Relation between the overlap densities of the three largest CMRJ
          clusters $q_i^{(\mathrm{CMRJ})}$, $i=1$, $2$, and $3$, and the two largest
          Houdayer clusters $q_1^{(\mathrm{H})}$, $q_2^{(\mathrm{H})}$ for
          $\antifrac=0.125$ and $L=32$. The inset shows the number of wrapping CMRJ
          clusters for three different system sizes. At $T_{\mathrm{CMRJ}}$ two
          wrapping clusters emerge, see also
          Fig.~\ref{fig:p0125_cmrj_percolation_plot}. Below $T_{\mathrm{f}}$, the
          number of wrapping clusters reduces from two to one.}
 \label{fig:p0125_cmrj_and_houdayer_clusters_plot}
\end{figure}

The main plot in Fig.~\ref{fig:p0125_cmrj_and_houdayer_clusters_plot} illustrates the
overlap properties of the largest CMRJ and Houdayer clusters. More precisely, it
shows the cluster densities multiplied by the sign of their overlap.  The replicas
are aligned so that the overlap of the largest cluster is always positive. Let us
first focus on Houdayer clusters. In the high-temperature phase, two percolating
Houdayer clusters exist, and they exhibit equal density but opposite overlap
signs. Consequently, the average overlap is zero. Below the ferromagnetic transition,
one cluster dominates the other and the overlap becomes non-zero. Turning to the CMRJ
clusters, their behavior is found to be quite analogous to that of the Houdayer
clusters, which is a consequence of CMRJ clusters being geometric subregions of
Houdayer clusters.  Following Eq.~\eqref{eq:occupation_probability_cmrj_clusters}, a
bond in the CMRJ construction can only be occupied when the overlap at the adjacent
lattice sites is identical. Consequently, as $T\to 0$, the two cluster types become
almost indistinguishable. The inset of
Fig.~\ref{fig:p0125_cmrj_and_houdayer_clusters_plot} shows the general behavior of
the number of wrapping CMRJ clusters as a function of temperature. At high
temperatures there is no percolation. At the CMRJ percolation transition,
$T_{\mathrm{CMRJ}}$, two percolating clusters emerge. Below the ferromagnetic phase
transition, $T_{\mathrm{f}}$, the number of wrapping CMRJ clusters decreases from two
to one.

\begin{table}[tb!]
  \caption{Critical quantities of the frustrated disordered ferromagnet for
    $\antifrac=0.125$. The results were obtained by finite-size scaling as
    described in Sec.~\ref{sec:observables_and_finite_size_scaling}. The
    ferromagnetic phase transition belongs to the disordered Ising universality
    class~\cite{HasenbuschEtAl2007TheUniversalityClassOf3DSiteDilutedAndBondDilutedIsingSystems}. The
    FKCK and the CMRJ percolation transitions are part of the random
    percolation universality class. For comparison, note that according to
    Ref.~\cite{WangEtAl2013BondAndSitePercolationInThreeDimensions} the values
    for random percolation are $\nu = 0.8764(12)$, $ \gamma/\nu = 2.046(6)$ and
    $\beta/\nu = 0.47705(15)$.}
  \begin{tabular}{ c  c  c  c }
    \toprule
    &magnetization& FKCK       & CMRJ \\ \hline
    $T_\mathrm{c}$&   3.2412(4)  &  4.02079(10) & 3.71523(24) \\
    $\nu         $&   0.687(9)   &  0.873(4)    & 0.875(6) \\
    $\gamma/\nu$&   1.972(10)    & 2.0451(24)   & 2.046(8) \\ 
    $\beta / \nu $  &   0.524(8) &   0.475(3)   & 0.474(8)\\ \botrule 	 
  \end{tabular}
\label{tab:critical_exponents_p0125}
\end{table}

\begin{figure}[tb!]
	\begin{center}
		\includegraphics{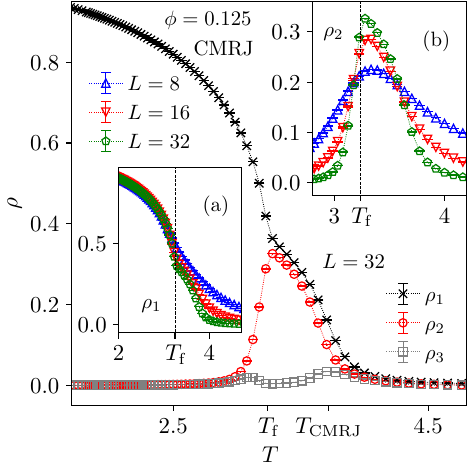}
	\end{center}
        \caption{The densities of the three largest CMRJ clusters as a function of
          temperature for $\antifrac=0.125$ and $L=32$. Inset (a) shows the behavior
          of the density $\rho_1$ of largest cluster for different system
          sizes. Inset (b) illustrates how the peak of the second largest cluster
          increases with system size and how it shifts towards the critical
          temperature of the ferromagnetic phase transition, $T_{\mathrm{f}}$.}
        \label{fig:p0125_three_largest_cmrj_clusters_plot}
\end{figure}

\begin{figure}[tb!]
	\begin{center}
		\includegraphics{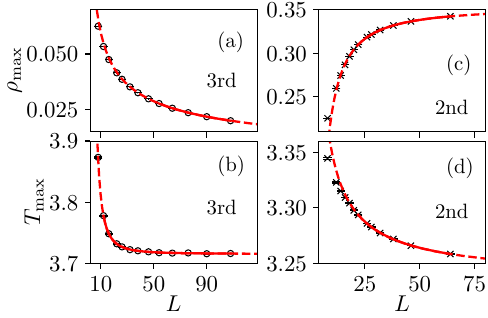}
	\end{center}
        \caption{Behavior of the peak densities of the third and the second largest
          CMRJ clusters as a function of system size for $\antifrac=0.125$. The red
          lines are power-law fits according to Eq.~\eqref{eq:peak_density_fitting}
          and Eq.~\eqref{eq:peak_temperature_fitting}, respectively. The dashed red
          lines are extrapolations of the fits. Panels (a) and (b) show the behavior
          of the peak of the third largest cluster, and panels (c) and (d) show the
          same relation for the second largest cluster.}
        \label{fig:p0125_peaks_of_cmrj_clusters_plot}
\end{figure}

Figure~\ref{fig:p0125_three_largest_cmrj_clusters_plot} illustrates the densities of
the three largest CMRJ clusters in more detail. In the vicinity of the CMRJ
percolation transition, $T_{\mathrm{CMRJ}}$, the third-largest cluster exhibits a
peak, which we denote as $\rho_{\mathrm{max}}^{(3)}$. As shown in panel (a) of
Fig.~\ref{fig:p0125_peaks_of_cmrj_clusters_plot}, this peak diminishes with
increasing system size. The decay can be approximated by a power law of the form
\begin{align}
\rho_{\mathrm{max}}^{(3)}(L)=a_{\rho_3}L^{-b_{\rho_3}}+c_{\rho_3},
\label{eq:peak_density_fitting}
\end{align} 
where the fitted parameters are $a_{\rho_3}=0.165(4)$, $b_{\rho_3}=0.426(12)$, and
$c_{\rho_3}=-0.0024(8)$. The smallest system size used in the fit is
$L_{\mathrm{min}}=54$, and the quality of the fit is
$Q_{\mathrm{fit}}=0.50$~\cite{Young2015EverythingYouWantedToKnowAboutDataAnalysis}. This
suggests that $\rho_{\mathrm{max}}^{(3)}$ approaches zero in the thermodynamic limit,
indicating that below the CMRJ percolation transition only the two largest clusters
have a nonzero density.

Additionally, the peak in the third-largest cluster size shifts in temperature with
increasing system size. This shift is also described by a power law,
\begin{align}
T_{\mathrm{max}}^{(3)}(L)=a_{T_3}L^{-b_{T_3}}+c_{T_3}.
\label{eq:peak_temperature_fitting}
\end{align}
Here, $c_{T_3}$ corresponds to the location of the peak in the thermodynamic
limit. The fit, shown in Fig.~\ref{fig:p0125_peaks_of_cmrj_clusters_plot} panel (b),
yields $a_{T_3}=14.8(1.9)$, $b_{T_3}=2.20(5)$, and $c_{T_3}=3.71572(26)$ with
$L_{\mathrm{min}}=12$ and $Q_{\mathrm{fit}}=0.94$. Notably, the value of $c_{T_3}$ is
consistent with the critical temperature of the CMRJ percolation transition,
$T_{\mathrm{CMRJ}}=3.71523(24)$.

The peak of the second-largest cluster follows a similar power law as the
third-largest cluster, though its density \emph{increases} with system size, see
inset (b) of Fig.~\ref{fig:p0125_three_largest_cmrj_clusters_plot}.
Figure~\ref{fig:p0125_peaks_of_cmrj_clusters_plot}, panels (c) and (d), shows that
the peak shifts towards lower temperatures.  Using the same fitting procedure as
above, the parameters are determined as $a_{\rho_2}=-2.45(10)$,
$b_{\rho_2}=1.309(14)$, $c_{\rho_2}=0.35271(23)$ with $L_{\mathrm{min}}=22$ and
$Q_{\mathrm{fit}}=0.27$, while the temperature-shift parameters are
$a_{T_2}=0.81(14)$, $b_{T_2}=0.85(7)$, and $c_{T_2}=3.235(4)$ with
$L_{\mathrm{min}}=22$ and $Q_{\mathrm{fit}}=0.80$.  This suggests that for
$L\to\infty$, the peak density of the second-largest cluster is approximately
$\rho_{\mathrm{max}}^{(2)}=0.35271(23)$, and that the peak location coincides with
the temperature of the ferromagnetic phase transition, since
$T_{\mathrm{max}}^{(2)}=3.235(4)$. Note that the shift exponent $b_{T_2}$ does not
appear to be compatible with $1/\nu \approx 1.145$ of the thermal transition.

\begin{figure}[tb!]
  \begin{center}
    \includegraphics{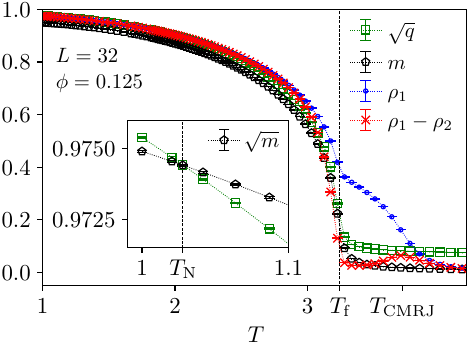}
  \end{center}
  \caption{Relation between the overlap, the magnetization, and the density of
    the two largest CMRJ clusters for $\antifrac=0.125$ and $L=32$ in the
    vicinity of the ferromagnetic phase transition. The density difference of
    the two largest clusters shows a similar behavior as the square root of the
    overlap, although the quantities are not identical. The inset illustrates
    that precisely on the Nishimori line $T_{\mathrm{N}}(0.125)=1.02779\ldots$
    the equality $q=m$ is
    satisfied~\cite{Nishimori1981InternalEnergySpecificHeatAndCorrelationFunctionOftheBondRandomIsingModel}.}
  \label{fig:p0125_overlap_and_cmrj_clusters_plot}
\end{figure}

As shown in the inset of
Fig.~\ref{fig:p0_magnetization_and_cmrj_largest_clusters_plot}, in the ordered phase
of the pure ferromagnet the square root of the overlap approximately equals the
density difference of the two largest CMRJ clusters, $\sqrt{q}=\rho_1-\rho_2$ for
$L\to\infty$, where $\rho_2=0$ in this case. Although this relation no longer applies
for $\antifrac>0$, Fig.~\ref{fig:p0125_overlap_and_cmrj_clusters_plot} shows that the
density difference of the two largest clusters still qualitatively reproduces the
behavior of the magnetization as well as the square-root of the overlap.

As a side note, the inset of Fig.~\ref{fig:p0125_overlap_and_cmrj_clusters_plot}
illustrates that on the Nishimori line the equality $m=q$ is satisfied. This follows
from the fact that on the Nishimori line the correlations of the magnetization and
the overlap
coincide~\cite{Nishimori1981InternalEnergySpecificHeatAndCorrelationFunctionOftheBondRandomIsingModel,HasenbuschEtAl2007MagneticGlassyMulticriticalBehaviorOfThe3DPMJIsingModel}.

\section{Clusters in the spin glass model}
\label{sec:clusters_in_the_standard_spin_glass}

\begin{figure}[t!]
  \begin{center}
    \includegraphics{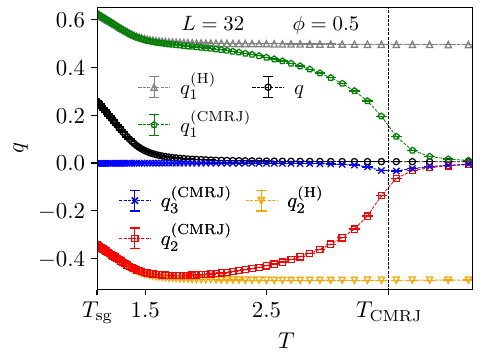}
  \end{center}
  \caption{Overlap densities of the largest Houdayer and CMRJ clusters for
    $\antifrac=0.5$ and system size $L=32$. At low temperatures CMRJ clusters are
    almost indistinguishable from Houdayer clusters.}
  \label{fig:p05_cmrj_and_houdayer_clusters_plot}
\end{figure}

For $\antifrac=0.5$, the low-temperature ordered state of the $\pm J$ random-bond
Ising model is of spin-glass type. At all temperatures, the system consists of two
large Ising clusters with opposite spin orientations, resulting in zero overall
magnetization. Furthermore, at high temperatures, there are two large Houdayer
clusters with opposite signs of the overlap. As the system undergoes a spin-glass
transition, a density difference emerges between these clusters, leading to a nonzero
total overlap, as shown in Fig.~\ref{fig:p05_cmrj_and_houdayer_clusters_plot}. This
transition belongs to the three-dimensional Ising spin-glass universality
class~\cite{HasenbuschEtAl2008CriticalBehaviorOf3DIsingSGModels}, with a critical
temperature
of~\cite{BaityJesiEtAl2013CriticalParametersOfThe3DIsingSG}~$T_{\mathrm{sg}}=1.1019(29)$.

A key distinction between this spin-glass transition and the previously discussed
ferromagnetic transitions is that the surface of the Houdayer clusters does not
exhibit critical behavior. The surface of the Houdayer clusters is connected to the
link overlap through the relation
\begin{align}
\widehat{Q}_\mathrm{l}=3N-2\widehat{Q}_{\mathrm{s}},
\end{align}
where $3N$ is the total number of bonds in the system, and $\widehat{Q}_{\mathrm{s}}$
is the number of bonds on the surface between regions with positive and negative
overlap~\cite{NewmanStein2007LocalVsGlobalVariablesForSpinGlasses}. The link overlap
is defined as
\begin{align}
    \widehat{Q}_\mathrm{l}=\sum_{\langle \bm{x}, \bm{y} \rangle} s_{\bm{x}}^{(1)} s_{\bm{y}}^{(1)} s_{\bm{x}}^{(2)} s_{\bm{y}}^{(2)}.
\end{align}
In the pure ferromagnet with $\phi=0$, the link overlap and the energy are related by
$q_\mathrm{l}=e_{\mathrm{I}}^2$, where
$q_\mathrm{l}=\langle \widehat{Q}_{\mathrm{l}}\rangle_{S_2}/(3N)$ and
$e_{\mathrm{I}}=\langle \widehat{H}_{\mathrm{I}} \rangle_S/(3N)$. Here,
$\widehat{H}_{\mathrm{I}} = - \sum_{\langle \bm{x}, \bm{y} \rangle } s_{\bm{x}}
s_{\bm{y}}$ denotes the Hamiltonian of a pure Ising ferromagnet. A similar
relationship between the link overlap and the energy also holds for spin glasses with
Gaussian couplings, see
Refs.~\cite{Contucci2003ReplicaEquivalenceInTheEaModel,Bray1980SomeObservationsOnTheMeanFieldTheoryOfSpinGlasses,MuensterWeigel2023ClusterPercolationInTheTwoDimensionalIsingSpinGlass}. Consequently,
at a ferromagnetic phase transition, the Houdayer cluster surfaces exhibit strong
fluctuations, as both the energy and the link overlap are critical quantities. In
contrast, at the spin-glass transition, the link overlap and the energy remain
non-singular. Note that the exponent $\alpha$, which describes the singular part of
the specific heat, is negative at the spin-glass
transition~\cite{BaityJesiEtAl2013CriticalParametersOfThe3DIsingSG}, $\alpha<0$.

For $\antifrac=0.5$, the FKCK percolation transition occurs at
$T_{\mathrm{FKCK}}=3.93410(7)$ and exhibits the same properties as a random
percolation transition with a single infinite cluster. In contrast, the CMRJ clusters
percolate at a lower temperature, $T_{\mathrm{CMRJ}}=3.50997(25)$. Both transitions
belong to the random-percolation universality class. The critical quantities for
these transitions are listed in Table~\ref{tab:critical_exponents_p05}. The results
were obtained by finite-size scaling, as described in
Sec.~\ref{sec:observables_and_finite_size_scaling}. In addition,
Table~\ref{tab:critical_exponents_p05} also includes critical exponents of the Ising
spin-glass transition taken from
Ref.~\cite{BaityJesiEtAl2013CriticalParametersOfThe3DIsingSG}.

\begin{figure}[tb!]
	\begin{center}
		\includegraphics{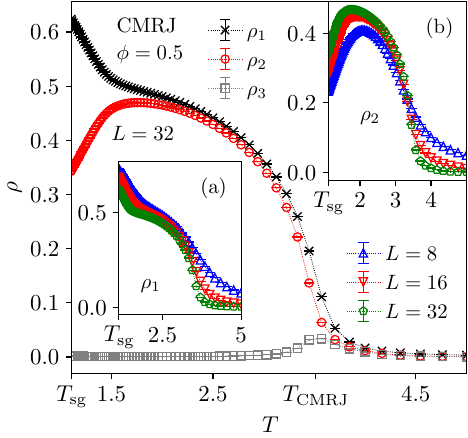}
	\end{center}
        \caption{The density of the three largest CMRJ clusters close to the CMRJ
          percolation transition for $\antifrac=0.5$ and system system size
          $L=32$. Insets (a) and (b) display the densities of the largest and
          second-largest clusters, respectively, for three different system sizes.}
        \label{fig:p05_three_largest_cmrj_clusters_plot}
\end{figure}

\begin{figure}[tb!]
	\begin{center}
		\includegraphics{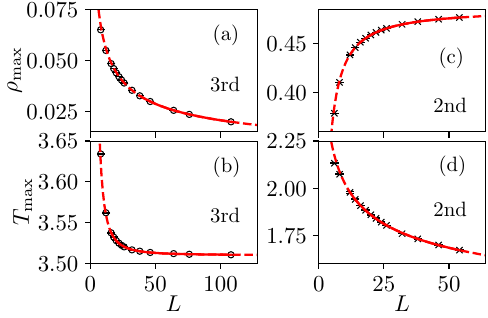}
	\end{center}
        \caption{Behavior of the peak densities of the third and the second largest
          CMRJ clusters as a function of system size for $\antifrac=0.5$. The solid
          red lines are power-law fits according to
          Eq.~\eqref{eq:peak_density_fitting} and
          Eq.~\eqref{eq:peak_temperature_fitting}, respectively. The dashed red lines
          are extrapolations of the fits. Panels (a) and (b) show the behavior of the
          peak of the third largest cluster, and panels (c) and (d) show the same
          relation for the second largest cluster.}
        \label{fig:p05_peaks_of_cmrj_clusters_plot}
\end{figure}

Figure~\ref{fig:p05_cmrj_and_houdayer_clusters_plot} illustrates that at the CMRJ
percolation transition, $T_{\mathrm{CMRJ}}$, two system-spanning clusters of equal
density emerge, which are subregions of the two Houdayer clusters. At the spin-glass
transition, these two clusters develop a density difference. In
Fig.~\ref{fig:p05_three_largest_cmrj_clusters_plot} we show the behavior of the three
largest CMRJ clusters in more detail. The third-largest cluster exhibits a peak close
to the CMRJ percolation transition. To analyze this peak as a function of system
size, the fitting approach from
Sec.~\ref{sec:clusters_in_the_frustrated_disordered_ferromagnet} is applied, using
Eqs.~\eqref{eq:peak_density_fitting} and~\eqref{eq:peak_temperature_fitting}. Panels
(a) and (b) of Fig.~\ref{fig:p05_peaks_of_cmrj_clusters_plot} show the peak behavior
of the density of the third-largest CMRJ cluster. The peak shifts towards the
critical temperature of the CMRJ percolation transition, with parameters
$a_{T_3} = 7.8(1.1)$, $b_{T_3} = 2.04(5)$, and $c_{T_3} = 3.51022(19)$ for
$L_{\mathrm{min}}=16$, with $Q_{\mathrm{fit}}=0.60$. Furthermore, the peak density
$\rho_{\mathrm{max}}^{(3)}$ decreases to zero as $L\to \infty$, suggesting that there
are only two macroscopic clusters below the CMRJ percolation transition. In this
case, the fitting parameters are $a_{\rho_3} = 0.1750(18)$, $b_{\rho_3} = 0.451(6)$,
and $c_{\rho_3} = -0.0012(4)$ with $L_{\mathrm{min}} = 38$ and
$Q_{\mathrm{fit}}=0.44$.

\begin{table}[b!]
  \caption{Critical quantities of the three-dimensional Ising spin glass at
    $\antifrac=0.5$.  The values of the spin-glass transition are taken from
    Ref.~\cite{BaityJesiEtAl2013CriticalParametersOfThe3DIsingSG}.  This transition
    belongs to the Ising spin-glass universality
    class~\cite{HasenbuschEtAl2008CriticalBehaviorOf3DIsingSGModels}.  The values of
    the FKCK and CMRJ percolation transitions were obtained by finite-size scaling as
    described in Sec.~\ref{sec:observables_and_finite_size_scaling}.  Both
    percolation transitions belong to the random percolation universality class.}
  \label{tab:critical_exponents_p05}
  \begin{tabular}{ c  c  c  c }
    \toprule
    &overlap (Ref.~\cite{BaityJesiEtAl2013CriticalParametersOfThe3DIsingSG}) & FKCK & CMRJ \\ \hline
    $T_\mathrm{c}$&1.1019(29)&3.93410(7)& 3.50997(25)\\
    $\nu         $&2.562(42)&0.873(4)& 0.878(7) \\
    $\gamma/\nu$&  2.39(9)   &  2.045(3) & 2.046(5)  \\   
    $\beta / \nu $&0.305(9)&  0.474(4) & 0.470(6) \\ \botrule 	
  \end{tabular}
\end{table}

Inset (b) of Fig.~\ref{fig:p05_three_largest_cmrj_clusters_plot} depicts how the
second-largest CMRJ cluster increases as a function of system size and shifts toward
lower temperatures. To study its properties, we again use the fitting procedure
according to Eqs.~\eqref{eq:peak_density_fitting}
and~\eqref{eq:peak_temperature_fitting}. The results are shown in panels (c) and (d)
of Fig.~\ref{fig:p05_peaks_of_cmrj_clusters_plot}, illustrating how the peak of the
second-largest CMRJ cluster shifts toward the spin-glass transition temperature, with
fitting parameters $a_{T_2} = 1.783(12)$, $b_{T_2} = 0.299(29)$, and
$c_{T_2} = 1.13(7)$ for $L_{\mathrm{min}}=12$ and
$Q_{\mathrm{fit}}=0.47$. Interestingly, the shift exponent $b_{T_2}$ is somewhat
comparable to the intrinsic shift exponent $1/\nu \approx 0.39$ of the spin-glass
transition. At its peak density, the second-largest cluster asymptotically contains
nearly half of the spin sites, since $c_{\rho_2} = 0.48582(9)$, with
$a_{\rho_2} = -0.709(8)$ and $b_{\rho_2}=1.092(5)$ for $L_{\mathrm{min}} = 16$ and
$Q_{\mathrm{fit}}=0.80$.

\begin{figure}[tb!]
	\begin{center}
		\includegraphics{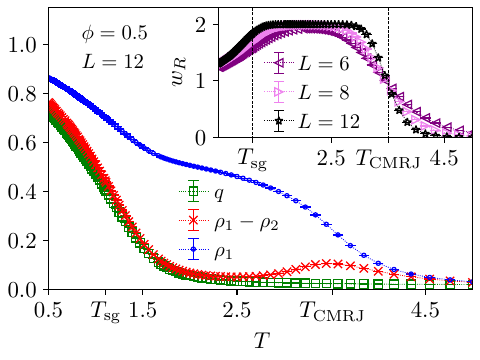}
	\end{center}
        \caption{Relation between the overlap and the density difference of the two
          largest CMRJ clusters for $\antifrac=0.5$ and system size $L=12$. The inset
          shows the number of wrapping CMRJ clusters for three different system
          sizes.}
        \label{fig:p05_overlap_and_cmrj_clusters_plot}
\end{figure}

Figure~\ref{fig:p05_overlap_and_cmrj_clusters_plot} demonstrates that there is a
direct relation between the density difference of the two largest clusters and the
overlap. The increase of this quantity below the spin-glass transition signals the
onset of spin-glass order and indicates the symmetry-broken
phase~\cite{MachtaNewmanStein2008ThePercolationSignatureOfTheSpinGlassTransition,MachtaNewmanStein2009APercolationTheoreticApproachToSGPhaseTransitions}. The
inset of Fig.~\ref{fig:p05_overlap_and_cmrj_clusters_plot} shows the number of
wrapping CMRJ clusters for three system sizes. At the CMRJ percolation transition,
$T_{\mathrm{CMRJ}}$, two wrapping clusters emerge. Below the spin-glass transition,
$T_{\mathrm{sg}}$, the number of wrapping clusters falls below two.

Notably, the stiffness of the vector spins $\tilde{s}_{\bm{x}}$ increases with the
temperature decreasing from $T_{\mathrm{CMRJ}}$ to $T_{\mathrm{sg}}$. To demonstrate
this, consider the Hamiltonian $\widehat{H}_{\bm{J}}^{(2)}$ of
Eq.~\eqref{eq:two_replica_hamiltonian}, which maintains both a global spin-reversal
symmetry $s_{\bm{x}} \to - s_{\bm{x}},~\forall \bm{x}$ and a vector-spin reversal
symmetry $\tilde{s}_{\bm{x}} \to - \tilde{s}_{\bm{x}},~\forall
\bm{x}$. Interestingly, the latter symmetry breaks at $T_{\mathrm{fr}}=2.045(23)$
when considering dynamics that preserve the site-wise overlap between two replicas,
$s_{\bm{x}}^{(1)}s_{\bm{x}}^{(2)}=q_{\bm{x}}=\mathrm{const},~\forall \bm{x}$. This
symmetry breaking suggests that the two-replica Hamiltonian
$\widehat{H}_{\bm{J}}^{(2)}$ exhibits a form of glassiness even before entering the
conventional spin-glass phase. Appendix~\ref{sec:conserved_overlap_transition}
provides a more detailed discussion of this conserved-overlap transition.

\section{Discussion}\label{sec:discussion}

In the present work we provided an extensive analysis of the relationship between
cluster percolation and thermal ordering in the three-dimensional $\pm J$ random-bond
Ising model. To develop a comprehensive understanding, both single-replica clusters,
such as Ising and FKCK variants, and two-replica clusters, such as the Houdayer and
CMRJ definitions, were studied. The latter two cluster types are particularly useful
because they are sensitive to the behavior of the overlap, which is the order
parameter of the spin-glass transition. The findings indicate that for all studied
fractions of antiferromagnetic bonds $0 \leq \antifrac \leq 0.5$, there are two giant
percolating Houdayer clusters of equal density at high temperatures. These two
clusters develop a density difference at the phase transition such that the average
overlap becomes nonzero. At ferromagnetic phase transitions, the surface of 
Houdayer clusters becomes critical, implying large fluctuations in this quantity. In
contrast, at the spin-glass transition, there is no singular behavior in the surface
of Houdayer clusters.

In the case of CMRJ clusters, the occupation probability decreases with increasing
temperatures. As a consequence, only small clusters exist at high temperatures, while
at low temperatures such clusters are almost identical to Houdayer clusters. In
the pure ferromagnet, when $\antifrac=0$, the CMRJ percolation transition maps onto
the ferromagnetic phase transition and shares the same critical behavior. For
$\antifrac>0$, the CMRJ percolation transition takes place at a higher temperature
than either the ferromagnetic or the spin-glass transition and belongs to the random
percolation universality class. Geometrically, CMRJ clusters are subregions of
Houdayer clusters. In agreement with this, below the percolation transition there are
two large CMRJ clusters of equal density that span the whole system such that the
number of wrapping clusters is two. At the ordering transition, the second-largest
cluster reaches its peak size before shrinking, causing the number of wrapping
clusters to drop below two. With increasing system size the locations of these peaks
of the second-largest clusters move to lower temperatures and converge to the
(ferromagnetic or spin-glass) ordering transitions, while the peak positions of the
third-largest clusters move towards higher temperatures and asymptotically reach the
corresponding percolation transitions.

The density difference between the two largest clusters qualitatively captures the
behavior of the overlap. For the Ising ferromagnet, the magnetization, the density
difference of the two largest clusters and the square-root of the overlap all
coincide in the ordered phase, and hence encapsulate the same scaling behavior. For
the disordered ferromagnet, the general correspondence of these quantities is
retained, but it is no longer quantitatively accurate. In contrast, for the spin
glass the magnetization is no longer meaningful, and the density difference of the
first and second clusters now corresponds to the overlap itself (instead of its
square-root), and hence to the order parameter of this transition.  This connection
allows the spin-glass transition to be identified as an imbalance between the two
largest CMRJ clusters. However, there is no direct equivalence between the overlap
correlation function and the connectivity function of the CMRJ
clusters~\cite{MachtaNewmanStein2008ThePercolationSignatureOfTheSpinGlassTransition,MachtaNewmanStein2009APercolationTheoreticApproachToSGPhaseTransitions}. The
main difficulty in identifying clusters that exhibit this property arises from the
presence of
frustration~\cite{CataudellaEtAl1994CriticalClustersAndEfficientDynamicsForFrustratedSpinModels}.

Beyond their physical significance, clusters are also used to construct powerful
non-local Monte Carlo
updates~\cite{Janke1998NonlocalMonteCarloAlgorithmsForStatisticalPhysicsApplications}. Cluster-based
algorithms using Houdayer or CMRJ clusters have proven effective in two-dimensional
spin
glasses~\cite{Houdayer2001ClusterMonteCarloAlgorithmFor2DimensionalSpinGlasses,Joerg2005ClusterMonteCarloAlgorithmsForDilutedSpinGlasses},
significantly accelerating equilibration in Monte Carlo simulations at low
temperatures. In three dimensions, however, the same algorithms provide only a modest
speedup. This is because the clusters percolate at temperatures higher than the
spin-glass transition, implying that the constructed structures are already
relatively stiff near the transition. This stiffness of the system is further
demonstrated by the conserved-overlap transition, which is described in more detail
in Appendix~\ref{sec:conserved_overlap_transition}. Therefore, to develop better
algorithms, it may be useful to identify other types of clusters or non-local
structures~\cite{KandelBenDomany1990ClusterDynamicsForFullyFrustratedSystems,CoddingtonHan1994GeneralizedClusterAlgorithmsForFrustratedSpinModels,CataudellaEtAl1996PercolationAndClusterMonteCarloDynamicsForSpinModels},
for instance by employing machine
learning~\cite{LiuEtAL2017SelfLearningMonteCarloMethod,McNaughtonEtAl2020BoostingMonteCarloSimulationsOfSpinGlassesUsingAutoregressiveNeuralNetworks}
or other simulation
techniques~\cite{Wang2017ExploringClusterMonteCarloUpdatesWithBoltzmannMachines,PeiDiVentra2022NonEquilibriumCriticalityAndEfficientExplorationOfGlassyLandscapesWithMemoryDynamics}. Additionally,
it may be useful to consider clusters that incorporate more than two
replicas~\cite{SwendsenWang1986ReplicaMonteCarloSimulationOfSpinGlasses,MuensterWeigel2023ClusterPercolationInTheTwoDimensionalIsingSpinGlass}. Some
consequences of this idea are explored in
Appendix~\ref{sec:nreplica_fkck_percolation}.

\appendix

\section{Correlations and Connectivity in CMRJ Clusters}
\label{sec:correlations_and_connectivity_in_cmrj_clusters}

We intend to show that for the case of the pure Ising ferromagnet CMRJ clusters
exhibit a relation between connectivity and correlations that is identical to that
observed for FKCK clusters. To this end, note that
\begin{align}
	\label{eq:vector_spin_and_spin_correlations}
	\left \langle
	\tilde{s}_{\bm{x}} \tilde{s}_{\bm{y}}
	 \right \rangle_{S_2} &= 
	 \sum_{\bm{S}^{(1)}, \bm{S}^{(2)}} \left( s_{\bm{x}}^{(1)}s_{\bm{y}}^{(1)}+s_{\bm{x}}^{(2)}s_{\bm{y}}^{(2)} \right) P_{S_2} \! \left(\bm{S}^{(1)},\bm{S}^{(2)} \right) \nonumber \\
	 &= 
	 2\sum_{\bm{S}} s_{\bm{x}} s_{\bm{y}} P_{S}\!\left(\bm{S} \right) 
	 = 2  \langle s_{\bm{x}} s_{\bm{y}}  \rangle_S, 
\end{align}
where 
\begin{align*}
P_S(\bm{S})= \frac{1}{Z} \exp\!\left( \sum_{\langle \bm{x}, \bm{y} \rangle }  s_{\bm{x}} s_{\bm{y}} /T \right), 
\end{align*}
and
\begin{align*}
  P_{S_2}(\bm{S}^{(1)},\bm{S}^{(2)})= \frac{1}{Z^2} \exp\left( \sum_{\langle \bm{x}, \bm{y} \rangle }  \tilde{s}_{\bm{x}} \tilde{s}_{\bm{y}} /T \right).
\end{align*}
Thus, according to Eq.~\eqref{eq:vector_spin_and_spin_correlations}, the
ferromagnetic phase transition described by the single-replica spins $s_{\bm{x}}$
maps onto the ferromagnetic phase transition of the vector spins
$\tilde{s}_{\bm{x}}$.

Now consider two lattice sites $\bm{x}$ and $\bm{y}$. If the corresponding spins
belong to the same CMRJ cluster, then $\tilde{s}_{ \bm{x} } \tilde{s}_{ \bm{y}}
=2$. If they belong to different CMRJ clusters,
$\tilde{s}_{ \bm{x} } \tilde{s}_{ \bm{y}}$ takes on a value in $\{-2,0,2\}$, where
due to spin-flip symmetry the cases $\pm 2$ occur with equal probabilities. From the
perspective of a Monte Carlo Markov chain based on CMRJ clusters, it is clear that
each CMRJ cluster can be flipped at random without violating detailed
balance~\cite{MuensterWeigel2023ClusterPercolationInTheTwoDimensionalIsingSpinGlass}. Flipping
a CMRJ cluster corresponds to reversing the sign of all vector spins in the cluster,
$\tilde{s}_{\bm{x}} \to -\tilde{s}_{\bm{x}},~\forall \bm{x}$ within the
cluster. Thus, on average, the product $\tilde{s}_{ \bm{x} } \tilde{s}_{ \bm{y}}$
equals $2$ if both spins belong to the same cluster and $0$ if they belong to
different clusters. From this observation, it follows that
\begin{align}
	\left \langle
	\tilde{s}_{\bm{x}} \tilde{s}_{\bm{y}}
	\right \rangle_{S_2} = 2 \,\mathrm{Prob}(\text{$\bm{x}$ and $\bm{y}$ are connected})
\end{align}
after averaging over all spin configurations and all configurations of occupied
bonds~\cite{Grimmett2004TheRandomClusterModel}. Combining this result with
Eq.~\eqref{eq:vector_spin_and_spin_correlations} shows that the connectivity function
of CMRJ clusters is identical to the spin--spin correlation function of a single
ferromagnet.
 
\section{Extrapolation ansatz for scaling corrections}
\label{sec:extrapolation_ansatz_for_scaling_corrections}

Consider the power-law scaling of an observable $O$ at the critical point, which
follows the general
form~\cite{AharonyAhlers1980UniversalRatiosAmongCorrectionToScalingAmplitudesAndEffectiveCriticalExponents,FerrenbergEtAl2018PushingTheLimitsOfMCSimulationForThe3DIsingModel}
\begin{align}
 \label{eq:powerlaw_corrections}
	O(L) &= c_0L^{\theta_0}+c_1L^{\theta_1}+c_2L^{\theta_2}+ \dots \\
	&=L^{\theta_0} \left(c_0+c_1L^{\tilde{\theta}_1}+c_2L^{ \tilde{\theta}_2}+ \dots  \right), \nonumber 
\end{align}
where we assume that $\theta_i>\theta_{i+1}$ with $i=0$, $1$, $2$, $3$, $\dots$, and
$\tilde{\theta}_j=\theta_j-\theta_0$, such that
$0>\tilde{\theta}_j>\tilde{\theta}_{j+1}$ with $j=1,2,3,\dots$. The terms
proportional to $L^{\theta_i}$ with $i \geq 0$ are scaling corrections that become
weaker in comparison to the leading order behavior for large system sizes because
$L^{\theta_i}/L^{\theta_0}=L^{\tilde{\theta}_i} \to 0$ for $L\to \infty$. By taking
the logarithm of Eq.~\eqref{eq:powerlaw_corrections} we get
\begin{align}
	\ln \left( O \right)=\theta_0 \ln(L)+\ln\left(c_0+c_1L^{\tilde{\theta}_1}+c_2L^{ \tilde{\theta}_2}+ \dots  \right).
 \nonumber
\end{align}
By substitution of $y_o=\ln \left( O \right)$ and $x= \ln(L)$ 
one obtains
\begin{align}
	y_o&=\theta_0 x+\ln\left(h(x) \right) ~~\text{where}   \\
	h(x)&=c_0+c_1e^{\tilde{\theta}_1x}+c_2e^{\tilde{\theta}_2x}+\dots \nonumber . 
\end{align}	 
Taking the derivative with respect to $x$ gives
\begin{align}
	\theta_{\mathrm{eff}}(x):=\frac{\mathrm{d} y_o}{\mathrm{d}x}=\theta_{0}+\frac{h'(x)}{h(x)}.
\end{align}
Because $\lim_{x\to \infty } \frac{\mathrm{d} y_o}{\mathrm{d}x} =\theta_{0}$ it follows that 
$h'(x)/h(x)\to 0$.  
If one only considers first-order scaling corrections $h_1(x)=c_0+c_1e^{\tilde{\theta}_1 x}$ 
the correction term becomes 
\begin{align}
\frac{h_1'(x)}{h_1(x)}&=\frac{\tilde{\theta}_1}{\frac{c_0}{c_1}e^{-\tilde{\theta}_1 x} + 1}.
\end{align}
Rewriting this as a function of $L$ leads to
\begin{align}
\frac{\tilde{\theta}_1}{\frac{c_0}{c_1}L^{-\tilde{\theta}_1} + 1}
\label{eq:extrapolation_ansatz}
\end{align}
and thus
$\theta_{\mathrm{eff}}\sim \theta_0+c
L^{\tilde{\theta}_1}$~\cite{AharonyAhlers1980UniversalRatiosAmongCorrectionToScalingAmplitudesAndEffectiveCriticalExponents}.
To numerically obtain the effective exponents, it is possible to fit a simple power
law $f(L)= a L^b$ to local regions of the data, where $a$ and
$b=\theta_{\mathrm{eff}}$ are fitting parameters. We denote such a local region which
extends from $L_{\mathrm{min}}$ to $L_{\mathrm{max}}$ and includes a fixed number of
system sizes, $W=\vert (L_{\mathrm{min}},\dots ,L_{\mathrm{max}} ) \vert$, as a fit
window.  The system sizes $L_k$ which are included in the fit windows originate from
a logarithmic scale $x_{k+1}= 2^{1/z} x_{k}$ where $k=1,2,3,4, \dots$. To be more
precise, $L_k$ are the rounded-up values of $x_k$ to the nearest even integer.

\begin{figure}[tb!]
	\begin{center}
		\includegraphics{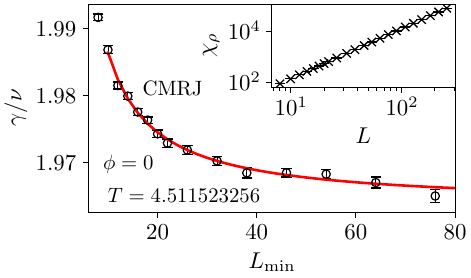}	
	\end{center}
        \caption{The main plot shows the dependency of the exponent $\gamma/\nu$ on the fit window for the average cluster size of the CMRJ clusters in the pure Ising ferromagnet ($\antifrac=0$). The red line is a fit of type $f(L_{\mathrm{min}}) = a_\theta +b_\theta/( c_\theta \, L_{\mathrm{min}}^{-b_\theta} + 1 )$ in accordance to Eq.~\eqref{eq:extrapolation_ansatz}, where $L_{\mathrm{min}}$ is the smallest system size of each fit window and $a_\theta=\gamma/\nu$ as well as $b_\theta$ and $c_\theta$ are fitting parameters. The result is $\gamma/\nu=1.9638(18)$ where the error bar is computed by bootstrapping as it is described in Ref.~\cite{Young2015EverythingYouWantedToKnowAboutDataAnalysis}. The inset shows the scaling of the average cluster size at $T=4.511\,523\,256$.}
        \label{fig:p0_cmrj_chi_exponent_plot}
\end{figure}

Figure~\ref{fig:p0_cmrj_chi_exponent_plot} illustrates the procedure for the average
cluster size of the CMRJ clusters at the critical temperature of the ferromagnetic
phase transition. The plot demonstrates how the critical exponent $\gamma/\nu$ is
approached by a power law as a function of $L_{\mathrm{min}}$.  The system sizes
reach from $L=8$ to $L=256$ and originate from a logarithmic scale with $z=4$. The
size of the fit window is $W=8$. Notably, a similar power-law extrapolation was also
performed in
Ref.~\cite{MuensterEtAl2021OrderingBehaviorOfTheTwoDimensionalIsingSGWithLongRangeCorrelatedDisorder}.

The extrapolation method is applied only for the pure ferromagnet with $\antifrac=0$.
The corresponding three values of $\gamma/\nu$, describing the singular behavior of
the magnetic susceptibility and the average cluster size of the FKCK and CMRJ
clusters, are listed in Table~\ref{tab:critical_exponents_p0}. For $\antifrac>0$, the
error due to corrections to scaling has been visually estimated from the convergence
behavior of $\gamma/\nu$ as a function of $L_{\mathrm{min}}$, i.e., without applying
a power-law extrapolation.

\section{The conserved overlap transition}
\label{sec:conserved_overlap_transition}

Here we show that the two-replica Hamiltonian $\widehat{H}_{\bm{J}}^{(2)}$, as
defined in Eq.~\eqref{eq:two_replica_hamiltonian}, undergoes a phase transition at a
higher temperature than the spin-glass transition under a specific dynamical
evolution. To explore this, assume that two replicas, $\bm{S}^{(1)}$ and
$\bm{S}^{(2)}$, are in equilibrium with respect to the Gibbs--Boltzmann distribution
for a given realization of disorder $\bm{J}$. At some point in the Monte Carlo
process, say at Monte Carlo time step $t_{\mathrm{fr}}$, the relative orientation of
the spin components is frozen, which means that
$s_{\bm{x}}^{(1)}(t_{\mathrm{fr}})s_{\bm{x}}^{(2)}(t_{\mathrm{fr}})=q_{\bm{x}}(t_{\mathrm{fr}})=\mathrm{const}$
from that time onward. This freezing procedure generates the subsidiary Hamiltonian
\begin{align}
\widehat{H}_{\bm{J}}^{(\mathrm{fr})} &= -\sum_{\langle \bm{x}, \bm{y} \rangle} J_{\bm{x}\bm{y}}^{(\mathrm{fr})} \tilde{v}_{\bm{x}} \tilde{v}_{\bm{y}},  
\label{eq:freezing_hamiltonian}
\end{align}
where
\begin{align*}
\tilde{v}_{\bm{x}} &= \frac{\pm 1}{\sqrt{2}}
\begin{pmatrix}
s_{\bm{x}}^{(1)}(t_{\mathrm{fr}}) \\
s_{\bm{x}}^{(2)}(t_{\mathrm{fr}})
\end{pmatrix}
\end{align*}
and $J_{\bm{x}\bm{y}}^{(\mathrm{fr})} = 2 J_{\bm{x}\bm{y}}$. The product
$\tilde{v}_{\bm{x}}\tilde{v}_{\bm{y}}=\pm [s_{\bm{x}}^{(1)}(t_{\mathrm{fr}})s_{\bm{y}}^{(1)}(t_{\mathrm{fr}})+s_{\bm{x}}^{(2)}(t_{\mathrm{fr}})s_{\bm{y}}^{(2)}(t_{\mathrm{fr}})]/2$
can take values $\pm 1$ or $0$. After freezing, the system evolves only within the
subspace of states where the site-wise overlap remains constant. To achieve this,
single-spin flip dynamics for the vector spins, i.e.,
$\tilde{v}_{\bm{x}} \to \pm \tilde{v}_{\bm{x}}$, is implemented alongside CMRJ
cluster updates~\cite{Joerg2005ClusterMonteCarloAlgorithmsForDilutedSpinGlasses},
which also preserve the
overlap~\cite{MuensterWeigel2023ClusterPercolationInTheTwoDimensionalIsingSpinGlass}.

The order parameter of the conserved-overlap transition is defined as $q_{\mathrm{fr}} := [\langle  \vert \widehat{q}_{\mathrm{fr}} \vert \rangle_V]_{J}$ with
\begin{align}
\widehat{q}_{\mathrm{fr}} = \frac{1}{N} \sum_{\bm{x}} \tilde{v}_{\bm{x}}^{(1)} \tilde{v}_{\bm{x}}^{(2)}.
\end{align} 
Here $\tilde{v}_{\bm{x}}^{(1)}$ and $\tilde{v}_{\bm{x}}^{(2)}$ belong to two replicas
of the system, i.e., $\tilde{v}_{\bm{x}}^{(1)} \in \bm{V}^{(1)}$ and
$\tilde{v}_{\bm{x}}^{(2)} \in \bm{V}^{(2)}$. These replicas are generated by making
two identical copies of the vector spins at $t_{\mathrm{fr}}$ and then evolving the
copies independently in time. This decorrelation of the two replicas requires most of
the simulation time. Furthermore, for the order parameter to function correctly, the
replicas must be properly aligned. At high temperatures, two giant Houdayer clusters
of equal density contain most of the spin sites. Thus, if the replicas $\bm{V}^{(1)}$
and $\bm{V}^{(2)}$ are misaligned, $\widehat{q}_{\mathrm{fr}}$ can vanish even below
the conserved-overlap transition, as long as the Houdayer clusters have equal
density. To avoid this, the observable is computed once, yielding the value
$\widehat{q}_{\mathrm{fr}}^{\,(a)}$. Next, the largest Houdayer cluster in replica
$\bm{V}^{(2)}$ is flipped and the observable is recomputed as
$\widehat{q}_{\mathrm{fr}}^{\,(b)}$. The final estimate is then given by
$\widehat{q}_{\mathrm{fr}}= \max( \vert
\widehat{q}_{\mathrm{fr}}^{\,(a)}\vert,\vert\widehat{q}_{\mathrm{fr}}^{\,(b)}\vert)$. Flipping
a Houdayer cluster means that all vector spins inside the cluster are reversed in
sign, $\tilde{v}_{\bm{x}}\to -\tilde{v}_{\bm{x}}$. This is a zero-energy
transformation because $\tilde{v}_{\bm{x}} \tilde{v}_{\bm{y}}=0$ at the cluster
surface, where $\bm{x}$ is a lattice site inside the cluster and $\bm{y}$ is a
lattice site outside the cluster.

\begin{figure}[tb!]
	\begin{center}
		\includegraphics{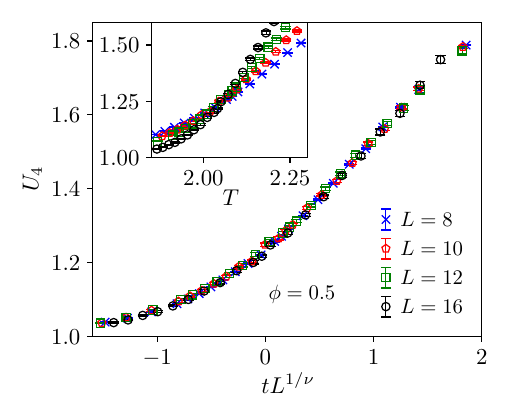}	
	\end{center}
        \caption{Data collapse of the $U_4$ parameter defined in
          Eq.~\eqref{eq:u4_parameter}. The inset shows the original data. The
          crossing of curves around $T_{\mathrm{fr}}=2.045(23)$ is clearly
          visible. The fraction of antiferromagnetic bonds is $\antifrac=0.5$.}
        \label{fig:p05vsg_u4_fss_plot}
\end{figure}

\begin{figure}[tb!]
	\begin{center}
		\includegraphics{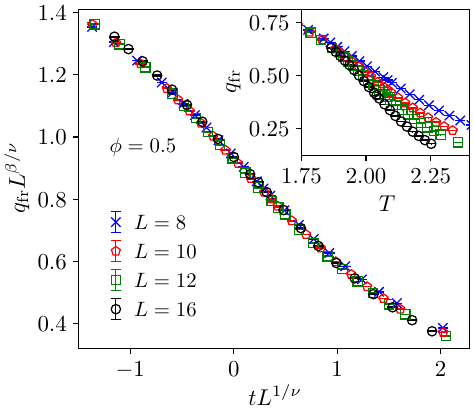}	
	\end{center}
        \caption{Data collapse of the conserved-overlap order parameter. The inset
          shows the original data. The fraction of antiferromagnetic bonds is
          $\antifrac=0.5$.}
        \label{fig:p05vsg_order_parameter_fss_plot}
\end{figure}

We investigate the conserved-overlap transition with the fraction of
antiferromagnetic bonds set to $\antifrac=0.5$. To analyze the critical behavior of
the transition, the order parameter and its higher moments are
studied. Figure~\ref{fig:p05vsg_u4_fss_plot} shows the dimensionless
quantity~\cite{Binder1981FiniteSizeScalingAnalysisOfIsingModelBlockDistributionFunctions,Binder1981CriticalPropertiesFromMonteCarloCoarseGrainingAndRenormalization}
\begin{align}
U_4=\frac{ \left[ \langle \widehat{q}_{\mathrm{fr}}^{\,4} \rangle_V \right]_{J} }{\left[ \langle \widehat{q}_{\mathrm{fr}}^{\,2} \rangle_V \right]_{J}^2} 
\label{eq:u4_parameter}
\end{align}
as a function of temperature. The crossing point of the data curves for different
system sizes indicates the location of the phase transition. By performing a data
collapse according to Eq.~\eqref{eq:fss_correlation_length}, the critical temperature
is obtained as $T_{\mathrm{fr}}=2.045(23)$ and the exponent as $\nu=1.00(8)$. The
relatively high transition temperature can be attributed to the fact that the
couplings of the Hamiltonian in Eq.~\eqref{eq:freezing_hamiltonian} are twice as
strong as the original couplings. Figure~\ref{fig:p05vsg_order_parameter_fss_plot}
shows a data collapse of the order parameter using
Eq.~\eqref{eq:fss_thermal_order_parameter}, yielding the exponent
$\beta/\nu=0.28(6)$. Additionally, the exponent $\gamma/\nu=2.33(10)$ is determined
by performing a fit according to Eq.~\eqref{eq:power_law_susceptibility} with system
sizes ranging from $L_{\mathrm{min}}=8$ to $L=16$. The critical exponents are similar
to those found for the phase transition between the ferromagnetic and spin-glass
phases in the three-dimensional $\pm J$ random-bond Ising
model~\cite{CeccarelliPelissettoVivari2011FerromagneticGlassyTransitionsIn3DIsingSGs}.

In general, it is difficult to obtain precise results for this transition, as long
simulation times are required to decorrelate the replicas $\bm{V}^{(1)}$ and
$\bm{V}^{(2)}$ after freezing at time step $t_{\mathrm{fr}}$. Each Monte Carlo time
step after freezing consists of one CMRJ cluster update and one sweep of single
vector-spin flips. The system sizes used range from $L=8$ to $L=16$. For $L=8$, up to
$181\,261$ disorder samples were generated with a decorrelation time of
$t_{\mathrm{d}}=120\,000$, whereas for $L=16$, up to $30\,875$ samples were generated
with $t_{\mathrm{d}}=1600\,000$.

\section{Multiple replica FKCK percolation}
\label{sec:nreplica_fkck_percolation}

In Sec.~\ref{sec:model_phase_diagram_and_cluster_definitions} we have introduced
Houdayer and CMRJ clusters, which are based on two replicas. Independently, Newman
and Stein proposed a method for defining clusters using two
replicas~\cite{NewmanStein2007ShortRangeSpinGlassesResultsAndSpeculations,MachtaNewmanStein2008ThePercolationSignatureOfTheSpinGlassTransition},
which can be directly extended to more than two
replicas~\cite{MuensterWeigel2023ClusterPercolationInTheTwoDimensionalIsingSpinGlass}. The
bond-occupation probability for these clusters with $K$ replicas is defined as
\begin{align}
p_{\bm{x}\bm{y}}^{(K)}=   
\begin{cases}
     (1-e^{-2/T} )^K&\text{if}~ J_{\bm{x}\bm{y}} \tilde{s}_{\bm{x}} \tilde{s}_{\bm{y}} = K \\
     0&\text{else}
\end{cases},
\end{align}
where
$\tilde{s}_{\bm{x}}\tilde{s}_{\bm{y}}=\sum_{k=1}^K s_{\bm{x}}^{(k)}s_{\bm{y}}^{(k)}$.
As a consequence, bonds can only be occupied if they are satisfied in all $K$
replicas simultaneously, i.e., $J_{\bm{x}\bm{y}} s_{\bm{x}}^{(k)} s_{\bm{y}}^{(k)}=1$
for all $k=1,2,\dots,K$. When $K=1$, the definition reduces to that of the standard
FKCK clusters given in Eq.~\eqref{eq:occupation_probability_fkck_clusters}.

Multiple-replica FKCK clusters were originally introduced as a mathematical tool to
prove broken symmetry in spin
glasses~\cite{NewmanStein2007ShortRangeSpinGlassesResultsAndSpeculations,MachtaNewmanStein2008ThePercolationSignatureOfTheSpinGlassTransition}. The
appearance of a unique infinite multiple-replica FKCK cluster with $K>1$ implies the
existence of broken symmetry in spin
glasses~\cite{NewmanStein2007ShortRangeSpinGlassesResultsAndSpeculations,MachtaNewmanStein2008ThePercolationSignatureOfTheSpinGlassTransition,MachtaNewmanStein2009APercolationTheoreticApproachToSGPhaseTransitions}. However,
their physical interpretation remains unclear.

The percolation transitions of these clusters were analyzed for different values of
$K$ and $\antifrac \in \{0,0.125,0.5\}$. Except for the FKCK clusters in the pure
Ising ferromagnet, all considered transitions belong to the random-percolation
universality class, as summarized in
Table~\ref{tab:crititical_quantities_k_replica_fkck}. For $\antifrac=0$ and $K>1$,
the percolation temperature $T_{\mathrm{c}}$ can be roughly estimated from the
bond-percolation threshold of the cubic
lattice~\cite{WangEtAl2013BondAndSitePercolationInThreeDimensions},
$p_{\mathrm{th}}=0.248\,811\,82(10)$, through the relation
$p_{\mathrm{th}} \approx (1-e^{-2/T_{\mathrm{c}}})^K$.

For $\antifrac=0.125$ and $K=2$, the clusters percolate at a temperature below the
ferromagnetic transition, while for $\antifrac=0.5$ they percolate above the
spin-glass temperature. For the considered numbers of replicas, there appears to be
no direct connection between the percolation of multiple-replica FKCK clusters and
ordering transitions. It would be interesting, however, to investigate whether a
choice of $K>2$ for the disordered cases leads to situations where the percolation
transition (nearly or exactly) coincides with the ordering transition.

\begin{table}
\caption{Critical quantities of some multiple replica FKCK percolation transitions for three different fractions of antiferromagnetic bonds. $K$ denotes to the number of replicas. The results were obtained by finite-size scaling as described in Sec.~\ref{sec:observables_and_finite_size_scaling}.}
\label{tab:crititical_quantities_k_replica_fkck}
	\begin{tabular}{ c  c  c  c }
    \toprule
          \multicolumn{4}{c}{$\antifrac=0$}  \\ \hline
		~~$K$~~ & $T_{\mathrm{c}}$  & $\nu$ & $\gamma/\nu$  \\ \hline
        1 & 4.511522(19)    &  0.630(4)   &  1.9630(21)    \\ 
		  2 & 2.80981(4)  & 0.876(7)  &  2.0454(25)   \\  
	    4  & 1.63019(4) & 0.877(5)   &  2.043(4) \\  \botrule
    \multicolumn{4}{c}{$\antifrac=0.125$}  \\ \hline
    $K$ & $T_{\mathrm{c}}$ & $\nu$ & $\gamma/\nu$ \\  \hline
      1 & 4.02079(10)   & 0.873(4)     & 2.0451(24)  \\ 
		 2 & 2.39619(18) & 0.874(8) &  2.043(4) \\  \botrule 
   \multicolumn{4}{c}{$\antifrac=0.5$}  \\ \hline
   $K$ & $T_{\mathrm{c}}$  & $\nu$ & $\gamma/\nu$    \\  \hline
    1  & 3.93410(7) & 0.873(4) & 2.045(3)   \\   
	2  & 1.7315(7) & 0.882(9)  & 2.040(10) \\ \botrule 
	\end{tabular}
\end{table}

\section{Simulation details}
\label{sec:simulation_details}

In our simulations, three types of Monte Carlo updates are employed: single-spin flip
updates, Swendsen--Wang cluster updates, and CMRJ cluster updates. The first two are
described, for example, in
Ref.~\cite{NewmanBarkema1999MonteCarloMethodsInStatisticalPhysics}. The CMRJ cluster
updates work as follows. First, the clusters are constructed according to the
probabilistic rule given in Eq.~\eqref{eq:occupation_probability_cmrj_clusters}. Then
each cluster is flipped with probability $0.5$. While CMRJ cluster updates satisfy
detailed balance with respect to the two-replica Gibbs--Boltzmann distribution, they
are not ergodic due to the conservation of overlap. Therefore, they need to be
augmented by ergodic updates such as single-spin flips or Swendsen--Wang cluster
updates. In addition to these single-replica moves, parallel tempering is implemented
with a chain of system copies simulated at different
temperatures~\cite{NewmanBarkema1999MonteCarloMethodsInStatisticalPhysics}. For the latter, we used between 20 and 40 temperatures, in general spaced uniformly in inverse temperature. In order to ensure sufficient acceptance rates for the swap moves of at least 10\%, additional temperature points were inserted when required. 

Depending on the fraction of antiferromagnetic bonds, different combinations of Monte
Carlo updates are used. For $\antifrac=0$, the Swendsen--Wang algorithm is employed
exclusively, such that each Monte Carlo time step consists of a single Swendsen--Wang
cluster update. As a criterion for equilibration, the Monte Carlo process is
initialized from both a ground-state configuration and a high-temperature
configuration, and the simulation is continued until both processes oscillate around
the same average value. Measurements are started after equilibration. To ensure that
the measurements are approximately uncorrelated, one sample is extracted only after
at least twice the integrated autocorrelation
time~\cite{Janke2008MonteCarloMethodsInClassicalStatisticalPhysics}. The system size
typically ranges from $L=16$ to $L=128$. For $L=16$, for example, 500\,000 samples
were generated, while for $L=128$ about 26\,000 samples were obtained to analyze the
ferromagnetic phase transition. Note that precisely at the critical temperature a
larger range of system sizes, from $L=8$ up to $L=256$, was simulated, as shown in
Fig.~\ref{fig:p0_cmrj_chi_exponent_plot}, with additional samples generated.

For $\antifrac>0$, Swendsen--Wang cluster updates combined with parallel tempering
were used to study the FKCK percolation transition, which occurs at relatively high
temperatures. At lower temperatures, one Monte Carlo time step consists of a sweep of
single-spin flips, a CMRJ cluster update, and a parallel tempering move. The
relaxation time is estimated from the time evolution of the link overlap, starting
from random initial spin configurations. The system is considered to be in
equilibrium when the link overlap reaches a stationary value in time, averaged over
disorder, see
Ref.~\cite{KatzgraberPalassiniYoung2001MonteCarloSimulationsOfSpinGlassesAtLowTemperatures}. For
the data analysis, at least 1000 disorder samples were generated for all system sizes
and lowest considered temperatures. The system size typically ranges from $L=16$ to
$L=90$ for studying the CMRJ percolation transition. For the data shown in
Fig.~\ref{fig:p0125_cmrj_percolation_plot}, for instance, the number of disorder
samples is 20\,597 for $L=32$ and 2886 for $L=108$.

\bibliography{percolation_3d_sg}
\end{document}